\documentclass[pra,aps,onecolumn,showpacs,showkeys,floatfix,superscriptaddress]{revtex4-1}

\usepackage[utf8]{inputenc}
\usepackage{color}
\usepackage{indentfirst}
\usepackage{latexsym,bm} 
\usepackage{amsmath,amssymb} 
\usepackage{graphicx}
\usepackage{multirow}
\usepackage{textcomp} 
\usepackage{hyperref}
\usepackage[outercaption]{sidecap}
\hypersetup{
	colorlinks=true
	linktoc=all
	}
\begin{document}

\title{\textit{MCTDH-X}: The multiconfigurational time-dependent Hartree method for indistinguishable particles software}

\author{Rui Lin}
\author{Paolo Molignini}
\affiliation{Institute for Theoretical Physics, ETH Zurich, 8093 Zurich, Switzerland}

\author{Luca Papariello}
\affiliation{Research Studio Data Science, RSA FG, 1090 Vienna, Austria}

\author{Marios C. Tsatsos}
\affiliation{ S\~{a}o Carlos Institute of Physics, University of S\~{a}o Paulo, PO Box 369, 13560-970, S\~{a}o Carlos, SP, Brazil}

\author{Camille L\'ev\^eque}
\affiliation{Vienna Center for Quantum Science and Technology, Atominstitut, TU Wien, Stadionallee 2, 1020 Vienna, Austria}
\affiliation{Wolfgang Pauli Institute c/o Faculty of Mathematics, University of Vienna, Oskar-Morgenstern Platz 1, 1090 Vienna, Austria}

\author{Storm E. Weiner}
\affiliation{Department of Physics,
	UC Berkeley,
	Berkeley, CA, United States
}

\author{Elke Fasshauer}
\affiliation{Department of Physics and Astronomy,
	Ny Munkegade 120,
	building 1520, 630,
	8000 Aarhus C, Denmark}

\author{R. Chitra}
\affiliation{Institute for Theoretical Physics, ETH Zurich, 8093 Zurich, Switzerland}

\author{Axel U. J. Lode}
\affiliation{Institute of Physics, Albert-Ludwig University of Freiburg, Hermann-Herder-Strasse 3, 79104 Freiburg, Germany}
\begin{abstract}
	
We introduce and describe the multiconfigurational time-depenent Hartree for indistinguishable particles (MCTDH-X) software, which is hosted, documented, and distributed at \mbox{\url{http://ultracold.org}}.
This powerful tool allows the investigation of ground state properties with time-independent Hamiltonians, and dynamics of interacting quantum many-body systems in different spatial dimensions. The MCTDH-X software is a set of programs
and scripts to compute, analyze, and visualize solutions for the
time-dependent and time-independent many-body Schr\"{o}dinger equation for indistinguishable quantum particles. As the MCTDH-X software represents a general solver for the Schr\"{o}dinger equation, it is applicable to a wide range of problems in the fields of atomic, optical, molecular physics as well as condensed matter systems. In particular, it can be used to study light-matter interactions, correlated dynamics of electrons in solid states, as well as some aspects related to quantum information and computing. The MCTDH-X software solves a set of non-linear coupled working equations based on the application of the variational principle to the Schr\"{o}dinger equation. These equations are obtained by using an ansatz for the many-body wavefunction that is a time-dependent expansion in a set of time-dependent, fully symmetrized bosonic (X=B) or fully anti-symmetrized fermionic (X=F) many-body basis states. It is the time-dependence of the basis set, that enables MCTDH-X to deal with quantum dynamics at a superior accuracy as
compared to, for instance, exact diagonalization approaches with a static basis, where the number of basis states necessary to capture the dynamics of the wavefunction typically grows rapidly with time. 

Herein, we give an introduction to the MCTDH-X software via an
easy-to-follow tutorial with a focus on accessibility. The illustrated exemplary problems are hosted at \mbox{\url{http://ultracold.org/tutorial}} and consider the
physics of a few interacting bosons or fermions in a double-well potential. We explore computationally the position-space and momentum-space density, the one-body reduced density matrix, Glauber correlation functions, phases, (dynamical) phase transitions as well as the imaging of the quantum systems. Although a few particles in a double well potential represent a minimal model system, we are able to demonstrate a rich variety of phenomena with it. We use the double well to illustrate the fermionization of bosonic particles, the crystallization of fermionic particles, characteristics of the superfluid and Mott-insulator quantum phases in Hubbard models, and even dynamical phase transitions.
We provide a complete set of input files and scripts to redo all computations in this paper at \url{http://ultracold.org/data/tutorial_input_files.zip}, accompanied by tutorial videos at \url{https://www.youtube.com/playlist?list=PLJIFUqmSeGBKxmLcCuk6dpILnni_uIFGu}. Our tutorial should guide the potential users to apply the MCTDH-X software also to more complex systems.

\end{abstract}
\maketitle

\section{Introduction}

The time-dependent many-body Schr\"{o}dinger equation (TDSE) is a fundamental equation at the heart of many different fields of science: quantum chemistry, condensed matter and atomic and molecular physics.
Exact solutions to the TDSE exist only for model systems, like the time-dependent harmonic interaction model~\cite{lode:12,lode:15,lode1:16,lode2:19}. Even for the time-independent many-body Schr\"{o}dinger equation (TISE), exact solutions are scarce ~\cite{girardeau,lieb:63,lieb2:63,mcguire:64,calogero:69,sutherland:71,schuck:01,yukalov:05}. 
These exact solutions, however, are in most cases not generalizable to current experiments or theoretical studies.
To obtain solutions to the TISE and TDSE, numerical methods as well as their implementation in software are therefore indispensable. Due to the fundamental nature of the problem, many methods have been put forward -- each with their advantages and shortcomings.

Examples of these numerical methods are matrix-product-state and density matrix renormalization group approaches, which employ a hierarchical partitioning of the many-body Hilbert space, and are particularly well-suited for one-dimensional lattice systems~\cite{schollwoeck:05,schollwoeck:11}. Meanwhile, mean-field approaches like the time-dependent Hartree-Fock~\cite{TDHF64} and the time-dependent Gross-Pitaevskii methods~\cite{gross:61,pitaevskii:61,TDMOMF} employ a drastically simplified approximation to the wavefunction of the state that ignores correlation effects.

Since the 1990s, when the multiconfigurational time-dependent Hartree approach~\cite{meyer:90,manthe:92,beck:00} (MCTDH) was first put forward, the method was applied very successfully in the field of theoretical chemistry, where systems involve coupled and distinguishable degrees of freedom. MCTDH enables the description of correlated wavefunctions; its ansatz for the wavefunction is a sum of all possible configurations of \textit{distinguishable} degrees of freedom or particles in a set of time-dependent variationally-optimized single-particle functions.
In 2003, the MCTDH for indistinguishable fermions~\cite{zanghellini:03,kato:04,caillat:05} (MCTDH-F) was formulated and in 2007 MCTDH for indistinguishable bosons~\cite{alexejsplit,alon:08} (MCTDH-B) followed. 
MCTDH-F and MCTDH-B can be formulated in a unified manner~\cite{alon.jcp:07}, where they share the same equations of motion; henceforth we will use the acronym MCTDH-X to refer to this unified prescription, where X is either X=F or X=B.

Other notable members of the MCTDH family of methods include the restricted active space (RAS-) and multilayer- (ML-)methods:
RAS-MCTDH-F~\cite{Miyagi:13}, RAS-MCTDH-B~\cite{Leveque:17,Leveque:18},
ML-MCTDH~\cite{wang:15,manthe:08,wang:03}, the ML-MCTDH in (optimized) second quantized representation~\cite{thoss:09} (\cite{manthe:17}), and the ML-MCTDH-X~\cite{schmelcher:17}. Here, the ``ML-'' prefix implies that a hierarchical format of the tensor representation of the many-body wavefunction is employed. For a review of multiconfigurational methods for the dynamics of indistinguishable particles including these multilayering and other methods, see Ref.~\cite{lode2:19}.

In this article, we provide an introduction to the software implementation of MCTDH-X hosted and distributed at \url{http://ultracold.org}~\cite{ultracold,lode1:16,lode2:16}.
In particular, MCTDH-F can be applied to describe the correlated dynamics of electrons in atoms and molecules~\cite{haxton:12,liao:17,Omiste2017_be,Omiste2018_neon,Omiste2018_neon,omiste:19} or to describe ultracold atomic fermions~\cite{lode1:16}.
MCTDH-B can be used to describe the many-body properties of ultracold atomic bosons with a focus on the phenomenon of fragmentation~\cite{spekkens:99,mueller:06}, where the reduced density matrix of the many-boson state attains several significant eigenvalues~\cite{streltsov:07,sakmann:09,sakmann.pra:10,sakmann:11,sakmann:14,lode2:17} and, as a result, quantum fluctuations are non-negligible~\cite{tsatsos:17,sakmann:16,lode:17,chatterjee:18,chatterjee2:19}.

Below, we provide a tutorial-type introduction to the MCTDH-X software with a focus on simplicity and instructiveness. The MCTDH-X software, however, can do way more than the examples we introduce below. The MCTDH-X software can deal with indistinguishable particles with internal degrees of freedom like spin~\cite{lode2:16}, indistinguishable particles placed in a high-finesse optical cavity~\cite{lode:17,lode.njp:18,molignini:18,lin2:19,lin:19}, indistinguishable particles with long-range dipolar interactions~\cite{chatterjee:15,chatterjee:18,chatterjee:19,chatterjee2:19} and Hubbard (lattice) models~\cite{lode2:16}. Moreover, the MCTDH-X software provides the possibility for an in-depth analysis of the computed solutions of the TISE and TDSE via full distribution functions~\cite{chatterjee2:19}, variances and quantum fluctuations~\cite{lode2:17,chatterjee2:19,tsatsos:17} of observables and correlation functions~\cite{chatterjee2:19,lode2:17,Lode2012,lode:15}; the MCTDH-X has been benchmarked against exact results~\cite{lode:12,lode:15}, verified against experimental predictions~\cite{tsatsos:17} and was recently reviewed in Ref.~\cite{lode2:19}.

The objective, workflow and usage of the MCTDH-X software is introduced in Sec.~\ref{sec:structure} and exemplified by a detailed tutorial in Sec.~\ref{sec:tutorial}, where ground states and dynamics of both bosons and fermions are inspected. Our focus is on introducing the usage of the software; details about the MCTDH-X theory are only complementarily discussed where necessary.
We conclude and summarize our work in Sec.~\ref{sec:conc}.

\section{Structure of the MCTDH-X Software}\label{sec:structure}

\subsection{Objective and main functionality}
The objective of the MCTDH-X software is to numerically solve the TISE or TDSE for a given many-body Hamiltonian, which describes $N$ interacting, indistinguishable bosons or fermions subject to a confining potential and to analyze the computed solutions. A general Hamiltonian has the form
\begin{equation}
\hat{H}=\sum_{i=1}^N \left[ -\frac{1}{2} \partial^2_{\mathbf{x}_i} + V(\mathbf{x}_i;t) \right] +  \sum^{N}_{i<j} W(\mathbf{x}_i,\mathbf{x}_j;t), \label{eq:hamiltonian}
\end{equation}
that can be either explicitly dependent on time or not. 
Here, $\mathbf{x}_i$ is the coordinate of the $i$-th particle, $-\frac{1}{2} \partial^2_{\mathbf{x}}$ is the kinetic energy operator, $V(\mathbf{x};t)$ is a general, possibly time-dependent, one-body potential and $W(\mathbf{x},\mathbf{x}';t)$ is a general, possibly time-dependent interparticle interaction operator. 
All the quantities are given in dimensionless units. The length scale $L$ can be chosen to appropriately represent the physical problem; the corresponding time and energy scales are then determined as $mL^2/\hbar$ and $\hbar^2/(mL^2)$, respectively. In particular, in the presence of a harmonic confinement potential, it is natural to choose the time scale as the inverse of the harmonic trapping frequency $\omega$, i.e., $L=\sqrt{\hbar/(m\omega)}$.

The time-independent many-particle Schr\"{o}dinger equation (TISE) corresponding to the Hamiltonian of Eq.~\eqref{eq:hamiltonian} is 
\begin{equation}
\hat{H} \vert \Psi_E \rangle = E \vert \Psi_E \rangle, \label{eq:TISE}
\end{equation}
while the time-dependent Schr\"odinger equation (TDSE) is
\begin{equation}
\hat{H} \vert \Psi (t) \rangle =  i \partial_t \vert \Psi(t) \rangle. \label{eq:TDSE}
\end{equation}
Note that $\hat{H}$ in the TISE needs to be a time-independent Hamiltonian.
In Eq.~\eqref{eq:TISE}, $\vert \Psi_E\rangle$ is an eigenstate of $\hat{H}$ with eigenvalue (energy) $E$. $\vert \Psi(t) \rangle$ stands for the solution of the TDSE at time $t$.  
 Technically, MCTDH-X is currently capable of accurately computing few lowest-in-energy eigenstates using Davidson or short iterative Lanczos routine from the Heidelberg MCTDH package~\cite{mctdh:package}.

The MCTDH-X theory~\cite{alon:08,alon.jcp:07} uses an ansatz for the wavefunction that is a time-dependent superposition of time-dependent many-body basis states:
\begin{eqnarray}
\vert \Psi(t) \rangle &=& \sum_{\vec{n}} C_{\vec{n}}(t) \vert \vec{n}; t \rangle; \;\; \vec{n}=\left(n_1,...,n_M\right)^T;\nonumber \\
\vert \vec{n}; t \rangle &=& \mathcal{N}  \prod_{i=1}^M \left[ \hat{b}_i^\dagger(t) \right]^{n_i} \vert \text{vac} \rangle; \qquad \phi_j(\mathbf{x};t)=\langle \mathbf{x} \vert \hat{b}_j (t) \vert 0 \rangle.  \label{eq:ansatz}
\end{eqnarray}
Here, the $C_{\vec{n}}(t)$ are referred to as coefficients, the $\vert \vec{n}; t \rangle$ as configurations, and the normalization factor is  $\mathcal{N}=\frac{1}{\sqrt{\prod_{i=1}^{M} n_i!}}$ for bosons and  $\mathcal{N}=1$ for fermions. Each configuration is a fully symmetric or fully anti-symmetric many-body basis state built from $M$ orthonormal time-dependent single-particle states,  or \emph{orbitals}, $\lbrace \phi_k(\mathbf{x},t); k=1,...,M \rbrace$. 
To fully specify the solution of the TISE or TDSE, the MCTDH-X software computes and stores the coefficients $C_{\vec{n}}(t)$ and the orbitals $\lbrace \phi_k(\mathbf{x},t); k=1,...,M \rbrace$ at times $t$ that are specified by the user. 

The set of equations of motion for the parameters in Eq.~\eqref{eq:ansatz} comprises a coupled set of first-order differential equations for time-dependent coefficients $C_{\vec{n}}(t)$ and non-linear integro-differential equations for the orbitals  $\phi_j(\mathbf{x};t)$. The details about these equations of motion and their derivation can be found, for instance, in Refs.~\cite{alon:08,alon.jcp:07,lode1:16,lode2:19}. 
The MCTDH-X software hosted at \url{http://ultracold.org} solves these equations of motion using the so-called constant mean-field integration scheme (see, for instance, Refs.~\cite{alon:08,beck:00}). The constant mean-field scheme features an adaptive time step for which the coefficients and the orbital's equations of motion are decoupled.

We note that the equation for the orbitals contains the inverse of the matrix elements of the one-body density matrix. In cases where the reduced one-body density matrix has zero eigenvalues and is not invertible, the orbital equations are therefore undefined and problematic. In almost all practical cases and, particularly, for the computations we present below, the regularization strategy documented in Ref.~\cite{beck:00} -- a posteriori adding negligibly small eigenvalues to make the inversion possible -- is sufficient. More elaborate schemes to improve or avoid this regularization have been developed~\cite{kloss:17,meyer:18}.

\subsection{Quantities of interest}\label{sec:quantities_interest}
Once the coefficients $C_{\vec{n}}(t)$ and the orbitals $\phi_k(\mathbf{x},t)$ are computed, the MCTDH-X software can analyze the solution and calculate several quantities of interest. These include, respectively, the real-space and momentum-space density distributions
	\begin{eqnarray}
	\rho(\mathbf{x})&=&\langle \Psi \vert \hat{\Psi}^\dagger(\mathbf{x}) \hat{\Psi}(\mathbf{x}) \vert \Psi \rangle/N, \label{eq:rhox} \\
	\tilde{\rho}(\mathbf{k})&=&\langle \Psi \vert \hat{\Psi}^\dagger(\mathbf{k}) \hat{\Psi}(\mathbf{k}) \vert \Psi \rangle/N,\label{eq:rhok}
	\end{eqnarray}
the Glauber one-body and two-body correlation functions,
\begin{eqnarray}\label{eq:corr1}
g^{(1)}(\mathbf{x},\mathbf{x}')&=& \frac{\langle \Psi \vert \hat{\Psi}^\dagger(\mathbf{x}) \hat{\Psi}(\mathbf{x}') \vert \Psi \rangle}{N\sqrt{\rho(\mathbf{x})\rho(\mathbf{x}')}}, \\
g^{(2)}(\mathbf{x},\mathbf{x}') &=& \frac{\langle \Psi \vert \hat{\Psi}^\dagger(\mathbf{x})  \hat{\Psi}^\dagger(\mathbf{x}') \hat{\Psi}(\mathbf{x}') \hat{\Psi}(\mathbf{x}) \vert \Psi \rangle}{N^2\rho(\mathbf{x})\rho(\mathbf{x}')},
\end{eqnarray}
and the natural orbitals $\phi_i^{(\mathrm{NO})}$ and the orbital occupations $\rho_i$, which, respectively, are the eigenfunctions and eigenvalues of the reduced one-body density matrix, 
\begin{equation}
\rho^{(1)}(\mathbf{x},\mathbf{x}')=\frac{1}{N}\langle \Psi \vert \hat{\Psi}^\dagger(\mathbf{x}') \hat{\Psi}(\mathbf{x}) \vert \Psi \rangle = \sum_i \rho_i \phi^{(\mathrm{NO}),*}_i(\mathbf{x}')\phi^{(\mathrm{NO})}_i(\mathbf{x}).\label{eq:RDM1}
\end{equation}
The natural orbitals are ranked in the order that the first orbital has the highest occupation while the last orbital has the lowest $\rho_1\ge\dots\ge\rho_M$. Another important quantity, the correlation order parameter (COP), is defined as the sum of the squares of the orbital occupations $\rho_i$~\cite{chatterjee:18,chatterjee2:19}.
\begin{eqnarray}\label{eq:fragmentation_OP}
\Delta=\sum_{i=1}^{M}\rho^2_i.
\end{eqnarray}
For the sake of brevity, we omit the dependence of quantities on time above and in the following, unless when imperative or instructive.

Even more importantly, the MCTDH-X software can simulate single-shot images~\cite{sakmann:16,lode:17,chatterjee:18,chatterjee2:19,tsatsos:17}. These single-shot images are the standard way to measure quantum many-body systems of ultracold atoms~\cite{bakr:09,buecker:09,sherson:10,smith:11}.
By simulating such single-shot images, the MCTDH-X software can reproduce the quantum measurements in the laboratory. 
In real space, a single-shot image can be obtained by drawing random positions
$(\tilde{\mathbf{x}}_1$, $\tilde{\mathbf{x}}_2  \dots \tilde{\mathbf{x}}_N)$ distributed according to the probability 
\begin{eqnarray}
P(\mathbf{x}_1,...,\mathbf{x}_N) = \left|\Psi(\mathbf{x}_1, \mathbf{x}_2, \dots, \mathbf{x}_N)\right|^2.
\end{eqnarray}
Due to the presence of quantum fluctuations and correlations, a single-shot image, which is distributed according to $\vert \Psi \vert^2$, can be drastically different than the density distributions $\rho(\mathbf{x})$, $\tilde{\rho}(\mathbf{k})$ in Eqs.~\eqref{eq:rhox} and \eqref{eq:rhok}. The deviation of single-shot images from the density distributions is especially significant when the particle number is small and/or the quantum correlations are large.

A large collection of single-shot images can be used to provide information on the system and particular its correlations and phases. Below, we consider a total number of $N_\mathrm{shot}$ images, with the value of the $i$-th image at position $x$ given by  $\mathcal{B}_i(\mathbf{x})$, and provide two types of single-shot analyses:
\begin{enumerate}
\item \emph{Single shots for particle correlations} between the two wells of a double-well potential: For each shot, the number of particles in one well, $n_i = \sum^\prime_{\mathbf{x}} \mathcal{B}_i(\mathbf{x})$, is calculated. Here, $\Sigma^\prime$ indicates that the summation is within a certain well (left or right).  We then calculate the probability of finding $n=n_i$ particles in the considered well among all single-shot images 
\begin{eqnarray}
P(n)= \frac{ N(n_i=n) }{N_\text{shot} }   ,    \quad n\in\{0,1,2,\dots,N\}  ,
\end{eqnarray}
where $N(n_i=n)$ denotes the total number of shots with $n_i=n$. We will show that the distribution of this probability depends on the correlations between particles. 
\item \emph{Quantum fluctuations} can also be extracted from single-shot images~\cite{tsatsos:17,lode:17,sakmann:16,chatterjee:18,chatterjee2:19}. To quantify the position-dependent quantum fluctuations of the particle number, we calculate the variance $\mathcal{V}(x)$ from single-shot simulations as:
\begin{eqnarray}\label{eq:var}
\mathcal{V}(\mathbf{x}) = \frac{1}{N_\text{shot}}\sum_{i=1}^{N_\text{shot}} \mathcal{B}_i^2(\mathbf{x}) - \left[\frac{1}{N_\text{shot}}\sum_{i=1}^{N_\text{shot}} \mathcal{B}_i(\mathbf{x})\right]^2.
\end{eqnarray}
\end{enumerate}
All these quantities above are chosen because of their accessibility in experiments and direct comparability to experimental results. For example, the momentum-space density distribution is accessible by time-of-flight measurements, the one-body particle correlations are accessible by thermodynamic quantities like kinetic energy~\cite{cocchi:17} and single-shot images are the standard way of measuring cold-atom systems~\cite{bakr:09,buecker:09,sherson:10,smith:11}. Other quantities of interest currently accessible by MCTDH-X include, for instance, the Glauber one-body and two-body correlation functions in momentum space~\cite{lode3:16} and many-body entropies of the system~\cite{Lode2015}.

\subsection{Workflow}
The MCTDH-X software mainly consists of two programs: 
\begin{enumerate}
\item The main program, \texttt{MCTDHX}, computes the numerical solution of the TISE or TDSE. 
\item The analysis program, \texttt{MCTDHX\_analysis}, analyzes the found solution. 
\end{enumerate}
Here and in the following, we use the \texttt{verbatim} font to refer to code, including executable commands, files, statements, and variables.

To set up a numerical task, the user modifies and chooses the parameters via the text input file \texttt{MCTDHX.inp}. A detailed description of the available options is given in the manual~\cite{manual} of the MCTDH-X software.
The file \texttt{Get\_1bodyPotential.F} is used to specify custom one-body potentials [$V(\mathbf{x}_i;t)$ in Eq.~\eqref{eq:hamiltonian}], while the file \texttt{Get\_Interparticle.F} allows for custom two-body potentials [$W(\mathbf{x}_i,\mathbf{x}_j;t)$ in Eq.~\eqref{eq:hamiltonian}]. 
For custom (initial) states, the \texttt{Get\_Initial\_Coefficients.F} and \texttt{Get\_Initial\_Orbitals.F} files can be used to specify the (initial) coefficients $C_{\vec{n}}(t=t_0)$ [cf. $C_{\vec{n}}(t)$ in Eq.~\eqref{eq:ansatz}] and orbitals $\lbrace \phi_k(\mathbf{x};t=t_0), k=1,...,M\rbrace$ [cf. $\phi_{j}(\mathbf{x};t)$ in Eq.~\eqref{eq:ansatz}], respectively.

\begin{figure}
	\includegraphics[scale=1]{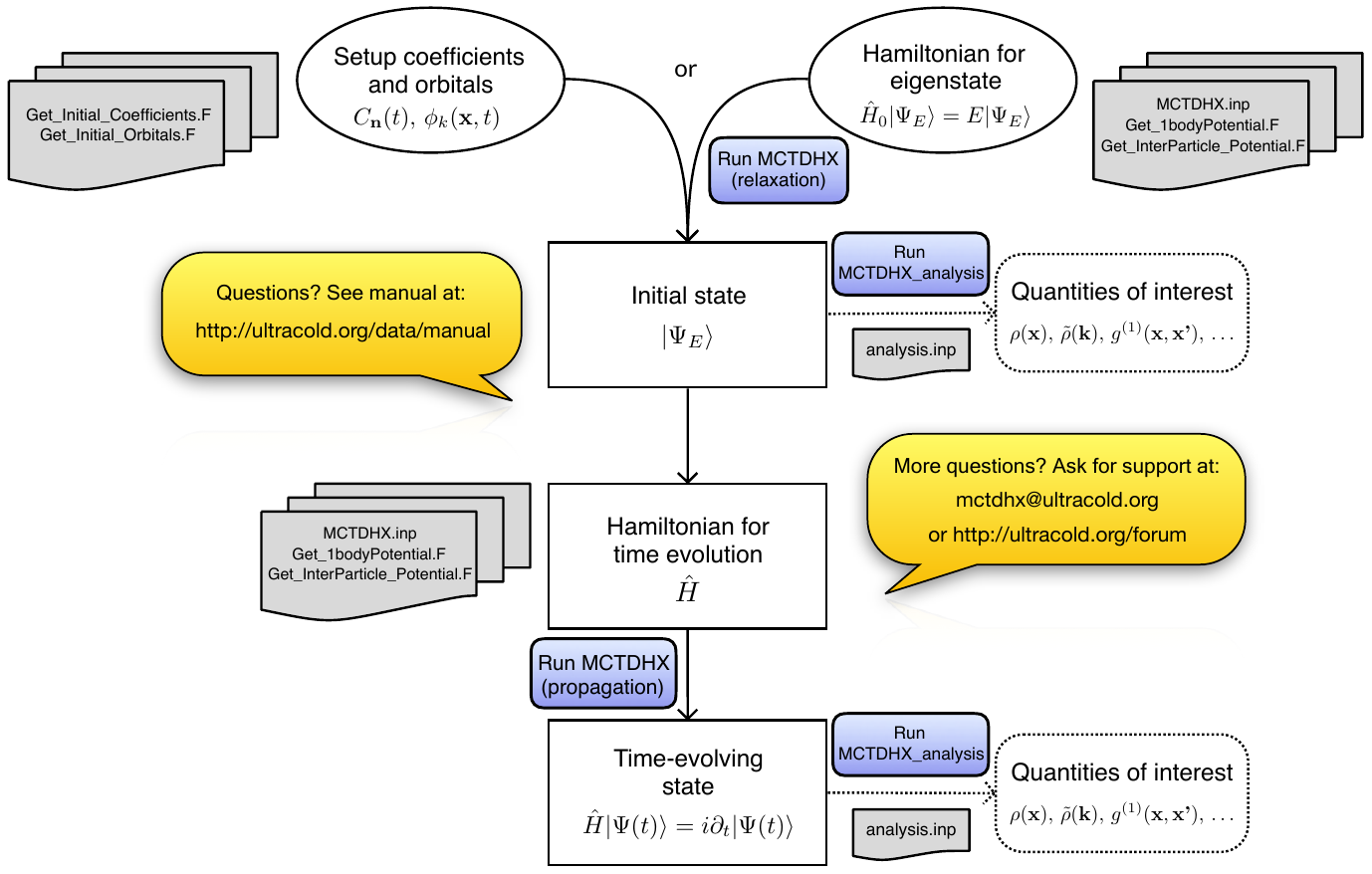}
	\caption{Workflow of the MCTDH-X software. }
	\label{fig:flowchart}
\end{figure}

The workflow and structure of the MCTDH-X software follows naturally from its main objectives to determine a numerical solution to the TISE or the TDSE and then to extract desired quantities of interest from the solution. This workflow can be summarized in the following steps, which are also visualized in Fig.~\ref{fig:flowchart}:\\[5mm]

\fbox{%
	\parbox{0.94\textwidth}{
\begin{enumerate}
    \item \textbf{Determine the initial state}. 
     \begin{enumerate}
     	\item Default: compute the ground state of some Hamiltonian $\hat{H}$ by running \texttt{MCTDHX} and configuring the numerical task (``relaxation mode'', Hamiltonian, integration procedure etc.) in the input file \texttt{MCTDHX.inp}.
    	\item Advanced: manually set the coefficients and orbitals that determine the initial wavefunction via the files \texttt{Get\_Initial\_Coefficients.F} and \texttt{Get\_Initial\_Orbitals.F}
     \end{enumerate}
		 
 \item \textbf{Analyze the initial state} by choosing the desired quantities of interest in \texttt{analysis.inp} and running the analysis program \texttt{MCTDHX\_analysis}
	\begin{enumerate}
		\item Default: call supplied visualization scripts or \texttt{gnuplot} to obtain plots or videos of the results
		\item Advanced: create custom visualizations or customize supplied visualization scripts
	\end{enumerate}
	
    \item \textbf{Compute the time-evolution of the initial state} with given Hamiltonian $\hat{H}$. The dynamics of the system is obtained by choosing the numerical task (``propagation mode'', Hamiltonian, integration procedure etc.) in the input file \texttt{MCTDHX.inp} and running \texttt{MCTDHX}.

    \item \textbf{Analyze the computed time-evolving state} by choosing the desired quantities of interest in \texttt{analysis.inp} and running the analysis program \texttt{MCTDHX\_analysis}. Visualization of the results as in step 2.(a) and 2.(b).
\end{enumerate}

}
}\\[5mm]

The scripts in the MCTDH-X software generally fall into two categories: either scripts to automate computations, i.e. parameter scans or scripts to visualize data in plots and videos. A series of tutorial videos illustrating the workflow step-by-step is available at \url{https://www.youtube.com/playlist?list=PLJIFUqmSeGBKxmLcCuk6dpILnni_uIFGu}. A discussion on the convergence, particularly in orbital number, can be found in the supplementary material~\cite{SUPPLMAT_TUTORIAL} (including Refs.~\cite{dutta:19,mistakidis:14,mistakidis:15,neuhaus-steinmetz:17,mistakidis:17,teichmann:09}).
For support and documentation, the website of the MCTDH-X software, \url{http://ultracold.org} ~\cite{ultracold}, the MCTDH-X manual~\cite{manual} and the email address \href{mailto:mctdhx@ultracold.org}{mctdhx@ultracold.org} are available.
Feature requests should be directed towards the support email address and/or be discussed on the web forum.

%



\section{Tutorial and application}\label{sec:tutorial}
As a tutorial example, we use the MCTDH-X software to study bosons and fermions in one-dimensional double-well one-body potentials. 
The double-well potential has a simple Hamiltonian, and is experimentally relevant~\cite{hofferberth:07b,betz:11,langen:15}.
Importantly, seen as a lattice with two sites, the double-well potential can be used as a demonstration on how we can use the MCTDH-X software to reveal the static and dynamical properties of the building blocks of lattice systems, i.e. periodic structures of potential wells. Such periodic structures are particularly interesting since they are commonly seen in nature and also synthesized in laboratories. Cold atom systems serve as one of such convenient platforms for the observation of quantum phase transitions~\cite{greiner:02}, since ultracold gases offer an astonishing flexibility -- for instance, the lattice depth and the inter-particle interaction strength can be varied almost at will. Choosing a minimal example like the double well reduces the computational difficulty and improves the clarity of our presentation. Nevetherless, some phenomena emerging in large lattices cannot be captured correctly, like for instance the Peierls transition, the SSH model~\cite{su:79,su:80} and more.

The TISE and TDSE for lattices are a cornerstone in many other complex physical systems studied in condensed matter physics, ultracold atoms and quantum gases, quantum computers, materials and more. Typically, the Hamiltonian for lattices (the above-discussed periodic potential) is approximated by the so-called Hubbard Hamiltonian (see \cite{SUPPLMAT_TUTORIAL} for details). This approximation uses a basis of static, potentially suboptimal site-localized functions, which are called Wannier functions. The Hubbard model features the the celebrated ``superfluid to Mott insulator'' transition as a result of the competition between the hopping between neighboring lattice sites and the on-site interaction~\cite{jaksch:98,greiner:02}. The double-well potential, though having only two lattice sites, also displays well this transition when varying the barrier between the two wells~\cite{jaaskelainen:05}.


A superfluid state of bosons is formed in a shallow lattice, where all particles in neighboring sites ``communicate'' and flow freely. It is characterized by quantum coherence between particles in distinct sites of the lattice. Such coherence results from the fact that all bosons occupy the same  single-particle state, i.e. $\vert \Psi \rangle \sim \vert N, 0, ... ; t \rangle $. 
In contrast, for deep lattice potentials, the particles are localized inside each site,  forming a ``frozen'' or insulating state analogous to the Mott insulator known from condensed matter physics. In a Mott-insulating state, all bosons in one potential well occupy the same single-particle state. 
Consider, for example, the wavefunction $\vert \Psi \rangle \sim \vert N/2, N/2, 0, ...; t \rangle$ for $N$ (here even) bosons in a double well. As a result, the Mott insulator state is characterized by the incoherence of particles between lattice sites.

The coherence and incoherence between sites in a lattice is also reflected in MCTDH-X simulations.
In MCTDH-X, to correctly simulate a Mott insulator state, each lattice site requires its own orbital.
If the number of orbitals $M$ is smaller than the number of lattice sites, two lattice sites have to ``share'' the same orbital, the emergence of Mott-type correlations between these two sites thus cannot be captured correctly.
With a sufficient number of orbitals, the coherence and incoherence between sites are accessible through quantities like the correlation functions [Eqs.~\eqref{eq:corr1}] or the reduced density matrix and its eigenvalues, which are straightforwardly available in simulations with the MCTDH-X software.

Even though the Bose-Hubbard model is a successful model, it is quite simplistic and cannot capture all the rich physics that might emerge.
MCTDH-X is able to capture the physics \emph{beyond} the Bose-Hubbard physics, by virtue of the used general basis set which is time-dependent, variationally-optimized and not necessarily localized at sites. These include a multi-band Bose-Hubbard model (see Ref.~\cite{SUPPLMAT_TUTORIAL}) and the fermionization of bosons. Specifically, situations where a Hubbard description breaks down have been found and the improved accuracy that MCTDH-X is necessary has been discussed in  Refs.~\cite{sakmann:09,sakmann:10,sakmann:11}.

In our examples, we follow and illustrate the workflow described in Sec.~\ref{sec:structure}. We first investigate the ground states and, subsequently the time evolution of one-dimensional setups with bosonic and fermionic particles, which are subject to double-well potential and repulsive interactions. The notation $\mathbf{x}$ will be replaced by $x$ in the following.

The double-well potential is modeled as a combination of an external harmonic confinement and a central Gaussian barrier. In \emph{natural} units, the MCTDH-X length $L=\sqrt{\hbar/m\omega}$ and energy $E=\hbar\omega$ scale are chosen according to an harmonic trapping frequency $\omega$. In these natural units, the double-well potential we consider is
\begin{eqnarray}\label{eq:double_well}
V_{\text{dw}}(x_i) = \frac{1}{2}x_i^2+E_\text{dw}\exp(-2x_i^2).
\end{eqnarray}
The two minima to the left and to the right of the Gaussian barrier correspond to two lattice sites, as visualized by the orange lines in Fig.~\ref{fig:double_well1}(a,d,g,j,m). The hopping strength of an analogous Hubbard model is mainly controlled by the barrier height, $E_\text{dw}$, while the on-site interaction is mainly determined by interaction strength, $g$ (see supplementary material \cite{SUPPLMAT_TUTORIAL} for details on the Hubbard model).

The interaction between the particles is chosen as contact interaction for bosons:
\begin{subequations}
\begin{eqnarray}
W_B(x_i,x_j) &=& g\delta(x_i-x_j),
\end{eqnarray}
and regularized Coulomb interaction for fermions:
\begin{eqnarray}
W_F(x_i,x_j) &=& \frac{g}{\sqrt{|x_i-x_j|^2+\alpha^2\exp(-\beta |x_i-x_j|)}}, \label{eq:fermion_interaction}
\end{eqnarray}
\end{subequations}
where $g>0$ is the repulsive interaction strength, $\delta(x)$ is the Dirac delta distribution and the parameters $\alpha=0.1$ and $\beta=100$ are used as in Refs.~\cite{lode1:16,bande:13}. These operators substitute $W$ in Eq.~\eqref{eq:hamiltonian} for the respective cases.

For illustration purposes and for the ease of computations, the number of particles is chosen to be small and thus far from the thermodynamic limit in our examples. The finite size of our systems renders the concept of phases and phase transitions less well-defined, because of the lack of non-analyticity in the ground state energy density during transition.
In the following, for the sake of simplicity and readability, we will use the terms ``phase'' and ``phase transition'' for discussing the properties of the quantum states being aware that they are only the finite-size precursors of the true quantum phases in the thermodynamic limit.
We will thus categorize states into different phases if they exhibit distinct behaviors of the various quantities of interest discussed in Sec.~\ref{sec:quantities_interest}.



All input files and scripts of this section are hosted at \url{http://ultracold.org/data/tutorial_input_files.zip} and a detailed description on how to do the computations is given in the supplementary material~\cite{SUPPLMAT_TUTORIAL} and a series of tutorial videos at \url{https://www.youtube.com/playlist?list=PLJIFUqmSeGBKxmLcCuk6dpILnni_uIFGu}.

\subsection{Ground state properties}

\begin{figure}[!]
	\includegraphics[width=0.9\textwidth]{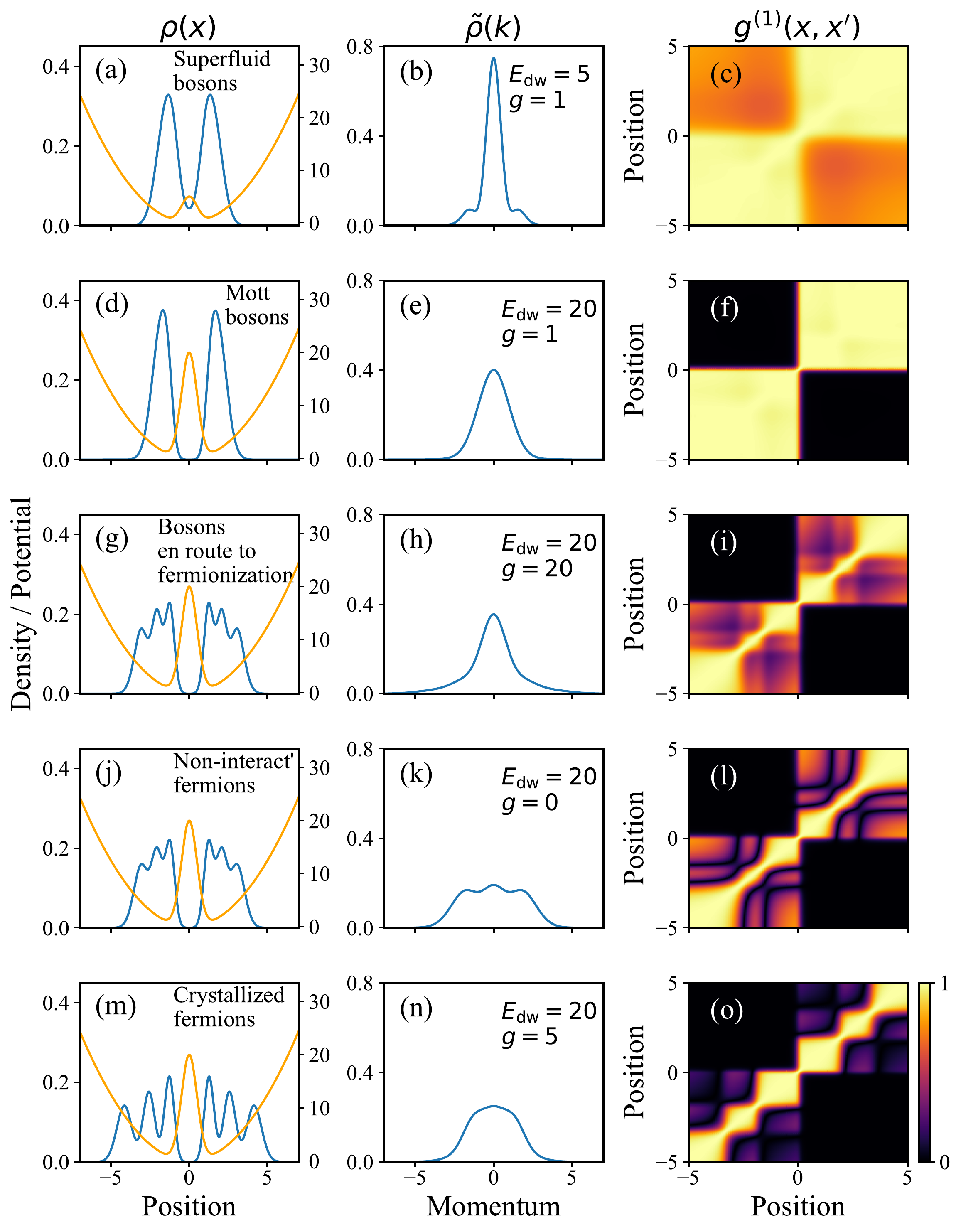}
	\caption{ The real-space density distribution $\rho(x)$  (first column blue lines, label on the left), potential $V(x)$ (first column orange lines, label on the right), momentum-space density distribution $\tilde{\rho}(k)$  (second column) and the one-body correlation function $g^{(1)}(x,x')$  (third column) of a superfluid bosonic state, (a-c), with $E_\text{dw}=5$, $g=1$ ($\rho_1\approx0.831$, $\rho_2\approx0.157$), a Mott insulating bosonic state, (d-f), with $E_\text{dw}=20$, $g=1$ ($\rho_1\approx0.497$, $\rho_2\approx0.493$), a bosonic state en route to fermionization, (g-i), with $E_\text{dw}=20$, $g=20$ ($\rho_1\approx\rho_2\approx0.246$, $\rho_3\approx\rho_4\approx0.152$, $\rho_5\approx\rho_6\approx0.102$), a non-interacting fermionic state, (j-l), with $E_\text{dw}=20$, $g=0$ ($\rho_1=\dots=\rho_6=0.167$) and a strongly-interacting crystallized fermionic state, (m-o), with $E_\text{dw}=20$, $g=5$ ($\rho_1\approx\rho_2\approx\rho_3\approx\rho_4\approx0.167$, $\rho_5\approx\rho_6\approx0.165$) . There are $N=6$ particles in all systems. The input files of the simulations are given as simulations \#1 to \#5 in Table~S1 of the supplementary material~\cite{SUPPLMAT_TUTORIAL}. } \label{fig:double_well1}
\end{figure}

We first solve the relevant TISE to find the ground state of $N=6$ bosons or fermions in the double-well potential of Eq.~\eqref{eq:double_well} by propagating the MCTDH-X coefficients $C_{\vec{n}}(t)$ and basis states $\phi_i(x;t)$ in imaginary time [cf. Eq.~\eqref{eq:ansatz}]. 
We vary the barrier heights $E_{\text{dw}}$ and interaction strengths $g$ to investigate where the superfluid-Mott insulator transition happens. The number of orbitals is chosen to be $M=10$. The density distributions in position space and momentum space and the one-body correlation function of the ground states in different potentials are shown in Fig.~\ref{fig:double_well1}.
 
A bosonic superfluid state and a bosonic Mott-insulator state are shown in Fig.~\ref{fig:double_well1}(a-c) and Fig.~\ref{fig:double_well1}(d-f), respectively. Compared to the Mott-insulator state, the superfluid state has two extra peaks in momentum space and a non-zero one-body correlation between the two wells $g^{(1)}(x,-x)>0$. The comparison between these two kinds of states has been investigated thoroughly with MCTDH-X~\cite{lode:17,roy:18,lin:19,lin2:19} and the results are consistent with other theoretical and experimental results~\cite{greiner:02,wessel:04,kato:08}. In this tutorial, we use the terms ``Mott insulating'' or ``Mott insulation'' to refer to situations where the one-body correlation function [Eq.~\eqref{eq:corr1}] has vanishing values in off-diagonal blocks, i.e., situations where $\vert g^{(1)}(x,x')\vert\approx 0$ is true for positions where $x$ and $x'$ are in distinct wells or at the position of distinct peaks of the one-body density, see Fig.~\ref{fig:double_well1}(f), for instance.

As the interaction between the bosons further increases, it induces a self-organized lattice-like structure of the density and correlations even within each of the double-well sites [Fig.~\ref{fig:double_well1}(g-i) for $g=20$]. This emergence of structure heralds the onset of so-called fermionization; the real-space densities of bosons with large contact interactions approach those of non-interacting fermions~\cite{girardeau:60}. 
The reason for the emergence of fermionization is that two bosons with infinitely large repulsive contact interactions, like fermions, cannot pass through each other in a one-dimensional system~\cite{girardeau:60}. Such similarities lay the foundation of useful tools like bosonization~\cite{senechal:99,giamarchi:03}.

However, fermionization and fermionic nature are driven by different mechanisms. This can be intuitively understood through their many-body wavefunctions. The wavefunction of a fermionized state of bosons, $\vert \Psi_B \rangle$, is still symmetric, while the wavefunction of a non-interacting fermionic state, $\vert \Psi_F \rangle$, is given by an anti-symmetric Slater determinant built from the first $N$ single-particle eigenstates. In position-space representation $\vert \Psi_B \vert^2 \equiv \vert \Psi_F\vert^2$ holds true in the fermionization limit; the wavefunctions of bosons and fermions, however, are completely different $\Psi_B(x_1,x_2,\dots,x_N) = \prod_{i<j}\mathrm{sgn}(x_i-x_j)\Psi_F(x_1,x_2,\dots,x_N)$~\cite{girardeau:60}. This difference is reflected in distinct features of the momentum distributions of bosons in the fermionization limit and non-interacting fermions [see Fig.~\ref{fig:double_well1}(b),(e),(h),(k),(n) and discussion below].

The density distribution at $g=20$ in Fig.~\ref{fig:double_well1}(g) is similar to the one of non-interacting fermions in Fig.~\ref{fig:double_well1}(j). The one-body correlation function of the fermions has a complicated pattern of significant and non-trivial correlations [Fig.~\ref{fig:double_well1}(l)]. The bosons have a slightly different correlation function than the fermions, but the distinct fermionic pattern is already visible. We thus refer to such states as ``en route to fermionization''. As discussed in Ref.~\cite{roy:18,bera:19} and the supplementary material, if an extremely large repulsive interaction and an adequate number of orbitals are used, the fermionization limit can be accurately captured by the MCTDH-X approach~\cite{SUPPLMAT_TUTORIAL}.

The difference between the bosons en route to fermionization and the fermions, however, shows most explicitly in momentum space. The momentum distribution for the bosons en route to fermionization [Fig.~\ref{fig:double_well1}(h)] has the same single-peak structure as the normal Mott insulating bosons [Fig.~\ref{fig:double_well1}(e)], where both widths and heights of the peaks are similar. This confirms the similarity between fermionized bosons and Bose-Einstein condensates in momentum space~\cite{girardeau:60}. 
In contrast, the fermionic state has three peaks in its momentum distribution due to the Pauli principle [Fig.~\ref{fig:double_well1}(k)]. These peaks correspond to the three fermions in each well since the two wells are Mott insulating [Fig.~\ref{fig:double_well1}(l)].

For long-range dipolar interactions crystallization emerges; for this case, bosons and fermions have been compared, for instance, in Ref.~\cite{deuretzbacher:10}.
A crystallized bosonic state and a crystallized fermionic state have similar real-space density distributions and they both have many contributing eigenvalues in the reduced one-body density matrix. This indicates that the long-range interaction dominates over the particle statistics.
We note, that crystallized bosons and fermionized bosons can be distinguished via the spread of their density matrices as a function of the interparticle interaction strength~\cite{bera:19}.

A fermionic state crystallizes in the presence of sufficiently strong long-range interactions $g=5$ [cf. Eq.~\eqref{eq:fermion_interaction}]. As expected, the repulsive interaction increases the distance between the fermions in real space, Fig.~\ref{fig:double_well1}(m). The interactions also have a pronounced impact in the momentum space distribution Fig.~\ref{fig:double_well1}(n) and the particle correlations Fig.~\ref{fig:double_well1}(o). The peaks in the real-space density distribution within each of the wells, which are induced by the Pauli exclusion principle in the absence of interaction [cf. Fig.~\ref{fig:double_well1}(j),(l)], become Mott insulating for crystallized fermions [cf. Fig.~\ref{fig:double_well1}(m),(o)].

\begin{figure}[!]
	\includegraphics[width=\textwidth]{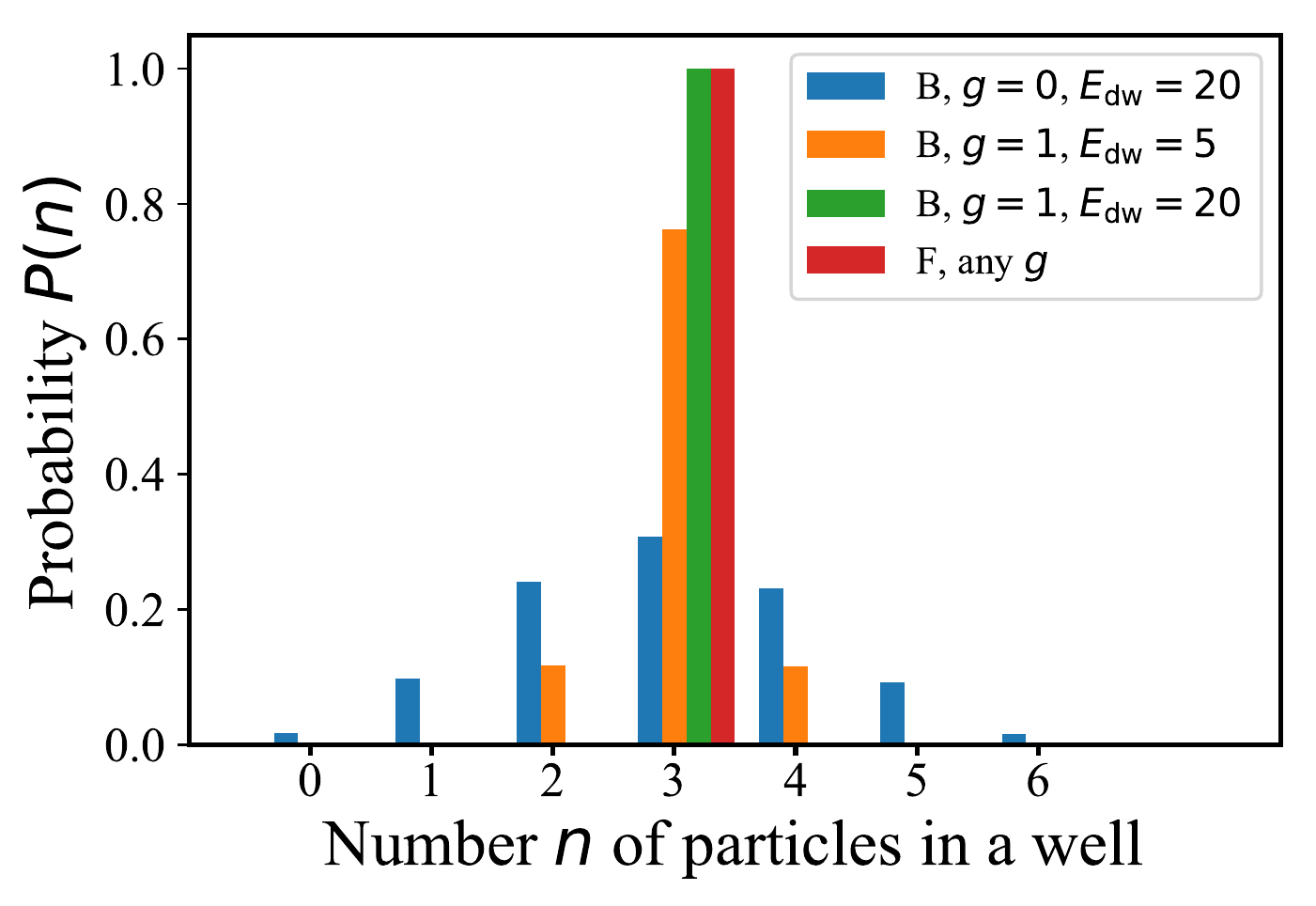}
	\caption{The probability $P(n)$ to find $n$ particles in the left well obtained from $10,000$ single-shot images (standard error $\mathcal{O}(1\%)$, not shown). For non-interacting bosons (blue), the probability follows a binomial distribution $P(0)=P(6)=1/64$, $P(1)=P(5)=6/64$, $P(2)=P(4)=15/64$ and $P(3) = 20/64$. For bosons with large interaction and fermions, the probability follows the prediction of the single-band Hubbard model and Pauli principles, respectively, both giving $P(3)=1$. The input files of the simulations are given as simulations \#1, \#2, \#5, \#6 in Table~S1 of the supplementary material~\cite{SUPPLMAT_TUTORIAL}. }  \label{fig:single_shot_gs}
\end{figure}

The correlations and fluctuations of particles can also be revealed by (simulations of) single-shot images~\cite{tsatsos:17,lode:17,sakmann:16,chatterjee:18,chatterjee2:19}. 
For an illustration of what can be extracted from simulated single-shots, we fix the barrier height $E_\text{dw}=20$ and compute bosonic and fermionic ground states for various interactions $g$. 
For every computed state, we generate $10,000$ simulated single-shots (adapting \texttt{analysis.inp} and running \texttt{MCTDHX\_analysis}). For each set of single-shot images, we calculate the frequency of $n=0,\dots,6$ particles being in the left well, $P(n)$, and show it in Fig.~\ref{fig:single_shot_gs}. For the fermions, each well contains exactly half of the particles regardless of the presence of interactions. This can be seen as a consequence of the Pauli principle. For the bosons, in the superfluid limit $g=0$, the system is in a coherent state [$g^{(1)}\sim 1$ for all $x,x'$ in Fig.~\ref{fig:double_well1}(c)]. In such a coherent state, the distributions of all bosons are independent from each other. As a result of this and the symmetry of the potential, $P(n)$ should be given by the binomial distribution $B(N,1/2) = \binom{N}{n}/2^N$ and this is confirmed by the single-shot results. As superfluidity gives way to Mott insulation, the distributions of different bosons become interdependent. Due to the repulsive interaction, the Hubbard model predicts that the particles tend to be distributed evenly in each well and $P(3)$ gradually increases while $P(n\neq3)$ gradually decreases.
For our MCTDH-X results in the Mott-insulator limit, $g=1$, $E_\mathrm{dw}=20$, the repulsion between particles becomes so strong that there are exactly three particles in each well $P(3)=1$ as predicted by the Hubbard model.

\begin{figure}[!]
	\includegraphics[width=\textwidth]{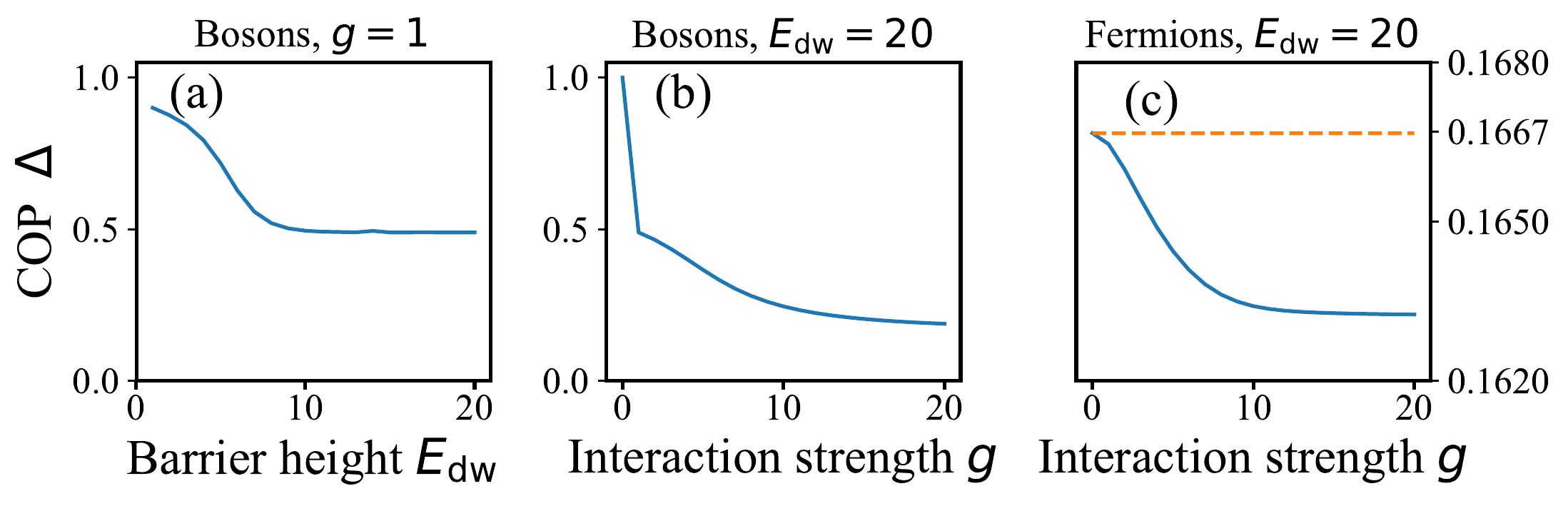}
	\caption{Correlation order parameter (COP) $\Delta$ [Eq.~\eqref{eq:fragmentation_OP}] as a function of barrier height $E_{\text{dw}}$ and interaction strength $g$ in double-well potentials. Six particles are used in all cases. In panel (a), the bosonic system transitions from the superfluid phase into the Mott insulator phase. In panel (b), the bosonic system goes from the superfluid phase at $g=0$ to the Mott insulator phase at small $g$ and finally to the crystallized phase at large $g$. In panels (a,b), $\Delta$ has a strong dependence on the phases. In panel (c), the fermionic system goes from the Mott insulating phase to the crystallized phase. $\Delta$ decreases only slightly when the fermions transition to the crystallized phase at larger values of $g$. The dashed orange line indicates $\Delta=1/6$ for non-interacting fermions. In all these simulations, the barrier is chosen as $E_\text{dw}=20$. The input files of the simulations are given in Table~S2 of the supplementary material~\cite{SUPPLMAT_TUTORIAL}. }  \label{fig:ground_occupation_OP}
\end{figure}

The behavior of the system in the superfluid, Mott-insulating and crystal phases can be summarized and represented by the correlation order parameter (COP) $\Delta$ [Eq.~\eqref{eq:fragmentation_OP}]~\cite{chatterjee:18,chatterjee2:19}.
The dependence of the COP on the barrier height $E_{\text{dw}}$ and interaction strength $g$ in bosonic and fermionic systems is shown in Fig.~\ref{fig:ground_occupation_OP}. In the bosonic system, the COP decreases from almost unity in the superfluid phase to $0.5$ in the Mott insulating phase, as expected for a fragmented state with $\rho_1\approx\rho_2\approx0.5$ [Fig.~\ref{fig:ground_occupation_OP}(a) and (b)]. As the interaction strength increases and the system is en route to fermionization, the COP drops further to $\Delta\approx0.2$. In the fermionic system, $\Delta$ is much less sensitive to the interaction strength. It only drops very slightly when the fermions crystallize [Fig.~\ref{fig:ground_occupation_OP}(c)]. The value $\Delta\approx0.167$ indicates each fermion occupies a single orbital, agreeing with the Pauli principle.

\subsection{Dynamical behavior}

Apart from solving the TISE for the ground state, MCTDH-X is also capable of solving the TDSE to capture the dynamics of a system as a reaction to a time-dependent Hamiltonian or to a quench of a parameter. As an example, we prepare the system in the ground state of a harmonic trap, $V_{\text{har}}(x)=\frac{1}{2} x^2$ and subsequently, we ramp up a barrier at $x=0$. We thus smoothly transform the harmonic trap into the double-well potential given in Eq.~\eqref{eq:double_well}. We use a linear ramp with a time scale $\tau$,
\begin{eqnarray}
E_\text{dw}(t) = \begin{cases}
E_\text{max}t/\tau,\quad &t\le\tau \\
E_\text{max},\quad &t>\tau.
\end{cases}
\end{eqnarray}
In the following, we compare the states that result for different $\tau$. 
There are two situations where the resulting state can be different from the ground state of the instantaneous Hamiltonian. \textbf{(i)} If the system undergoes a first-order phase transition -- while the parameters of the Hamiltonian change smoothly, the system can get stuck in an excited state of the final Hamiltonian in the absence of a strong perturbation.  \textbf{(ii)} With a fast and non-adiabatic change of parameters in the Hamiltonian, excitations are inevitably triggered. Such a quench thus may result in a finite occupation of a large number of eigenstates of the final Hamiltonian. To investigate these two scenarios, we now discuss two examples where the system contains $N=5$ or $N=6$ weakly-interacting bosons or fermions.

\begin{figure}[!]
	\includegraphics[width=\textwidth]{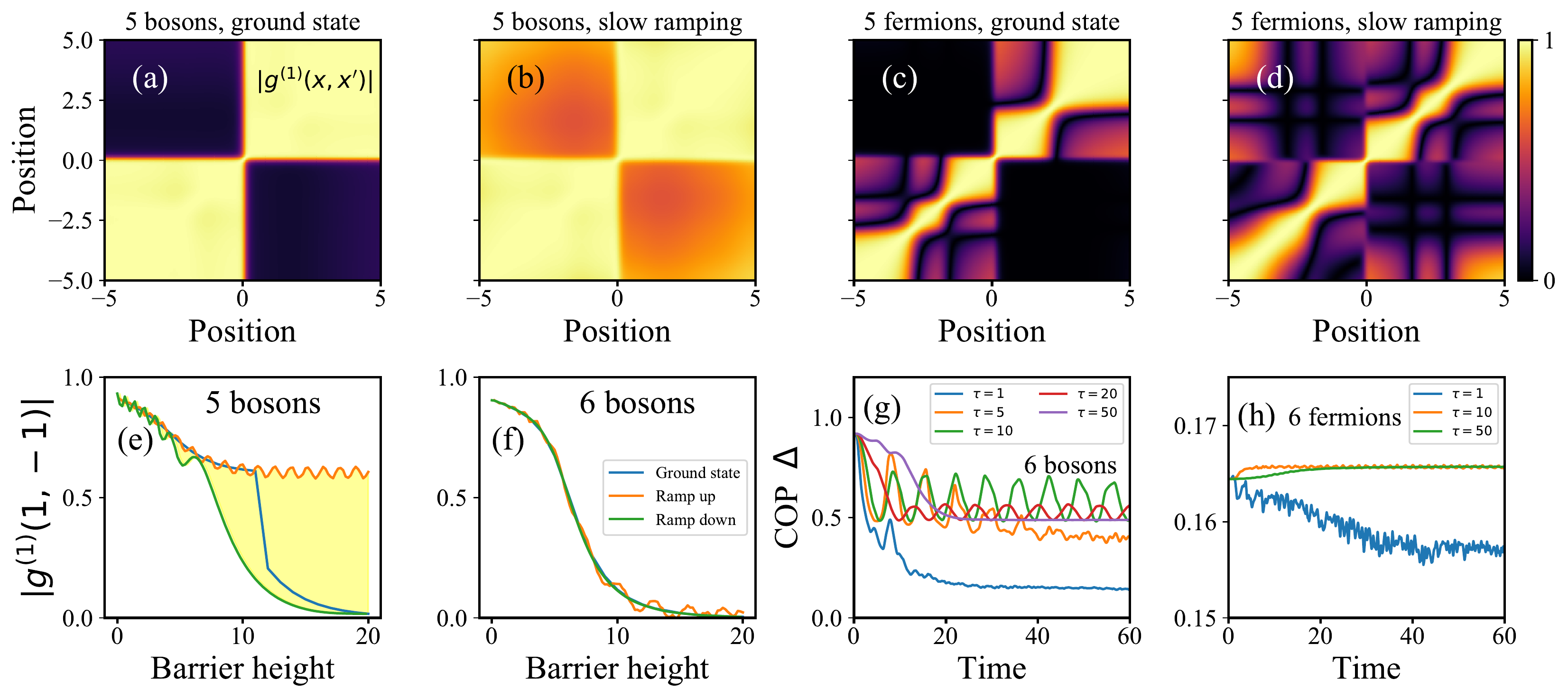}
	\caption{(a-d) One-body correlation functions $\vert g^{(1)}(x,x') \vert$ of $N=5$ weakly-interacting ($g=1$) bosons (a,b) and $N=5$ weakly-interacting ($g=2$) fermions (c,d). 	
	In (a,c) the ground states and in (b,d) slowly evolving states are shown. (e,f) The one-body correlations between the two wells $\vert g^{(1)}(1,-1)\vert$ as a function of barrier height (blue) in the ground state, when the barrier is ramping up (orange) and ramping down (green) slowly, for a system with (e) $N=5$ and (f) $N=6$ bosons. Hysteresis is clearly seen and marked in yellow in panel (e) but absent in panel (f). The ramping time is chosen as $\tau=100$ for all cases in panels (b), (d), (e) and (f), and the correlation functions at $t=100$ are shown in panels (b) and (d).
	The input files of the dynamical simulations are given as simulations \#13 to \#17 in Table~S3 of the supplementary material~\cite{SUPPLMAT_TUTORIAL}.
	(g,h) Correlation order parameter (COP) $\Delta$ [Eq.~\eqref{eq:fragmentation_OP}] of (g) $N=6$ bosons and (h) $N=6$ fermions as a function of time with different ramping times $\tau$. The dynamical behavior exhibits two different regimes, one for $\tau\lesssim10$ and the other for $\tau\gtrsim10$ for bosons and fermions alike, implying a dynamical phase transition at $\tau\simeq10$. The input files of the simulations are given as simulations \#20 and \#21  in Table~S4 of the supplementary material~\cite{SUPPLMAT_TUTORIAL}.} \label{fig:bosons_hysteresis}
\end{figure}

A first-order transition between a superfluid and a Mott insulator has already been observed in the presence of a three-body interaction~\cite{savavi:12} in a bosonic system. Such a first-order transition is also observed in our simulations even without three-body interactions, when the total number of bosons or fermions is odd ($N=5$ in our simulations). 
There is a large residual correlation $\vert g^{(1)}(x,-x)\vert \approx0.6$ in the time-evolving state -- even in the slow evolution limit with ramping time $\tau=100$ [Fig.~\ref{fig:bosons_hysteresis}(b,d)] for both, the bosonic and the fermionic cases, compared to the ground state where the one-body correlations between the two wells vanish completely
[$x,x'$ where $\vert g^{(1)} \vert \approx 0$ in Fig.~\ref{fig:bosons_hysteresis}(a,c)]. 

The interaction between particles is an essential ingredient in this first-order transition. It lifts the degeneracy between the Mott-insulating state and the state with large residual correlations. However, as the barrier of the double well increases, one of the particles has no preference for entering either of the two wells and thus straddles across both wells. This straddling particle builds up the correlations between the two wells that we observe.

To justify the claim of a first-order transition, we investigate the hysteretical behavior of one-body correlations $\vert g^{(1)}(1,-1)\vert$ for $N=5$ bosons, as shown in Fig.~\ref{fig:bosons_hysteresis}(e). In the ground state, there is clearly a jump at roughly $E_\text{dw}\approx11$. To study the dynamical effects, we choose a ground state in the superfluid limit, $E_\text{dw}=0$ and a ground state in the Mott-insulator limit, $E_\text{dw}=20$. We propagate them in ``opposite directions'' across the phase boundary by slowly ramping up (down) the barrier from $E_\text{dw}=0$ to $E_\text{dw}=20$ ($E_\text{dw}=20$ to $E_\text{dw}=0$) in a time of $\tau=100$. It is clear that there is a hysteresis in the particle correlations. Strikingly, the hysteresis covers an infinite area [see the yellow marked region in Fig.~\ref{fig:bosons_hysteresis}(e)], as the residual correlations will never vanish in the Mott limit. 
The jump in the orbital occupation and the hysteresis is not seen with an even number of particles [Fig.~\ref{fig:bosons_hysteresis}(f)]. In this even-number case, it is thus a second-order phase transition.

\begin{figure}[!]
	\includegraphics[width=0.8\textwidth]{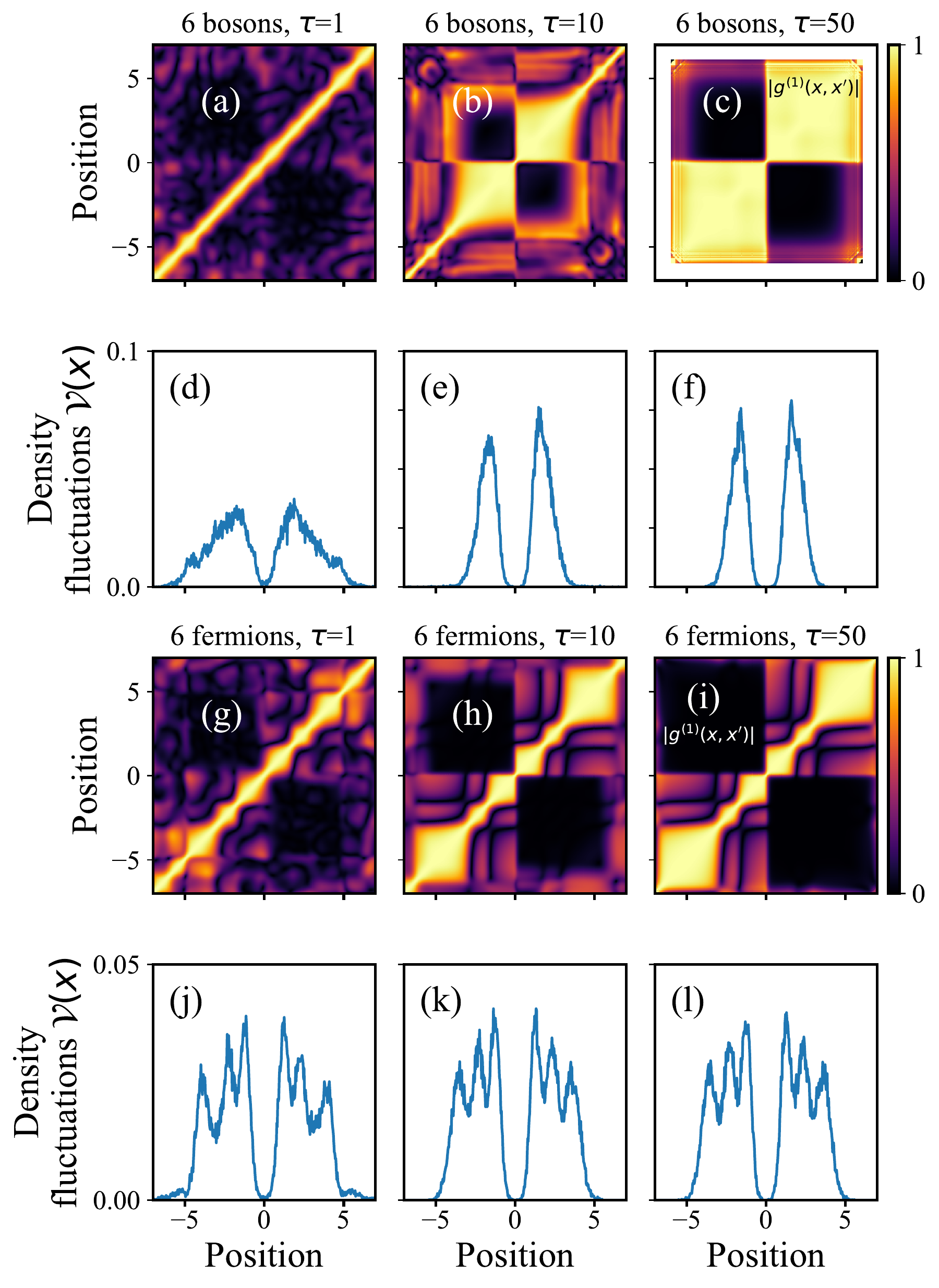}
	\caption{One-body correlation functions $\vert g^{(1)}(x,x') \vert$ (a-c) and variance $\mathcal{V}(x)$ over $10,000$ single-shot experiments (d-f)  of $N=6$ bosons at $t=80$ with different ramping times $\tau=1, 10, 50$. At fast ramping, $\tau=1$, the correlations $\vert g^{(1)}\vert$ are fluctuative. In the bosonic system, the distribution of the single-shot variance depends significantly on the ramping rate. The same quantities as in (a-f), but for fermions (g-l). The one-body correlation functions have a quite similar structure to those of the bosons [compare panels (g-i) and (a-c)]. 
	Unlike for the bosonic case, the density fluctuations quantified via the variance $\mathcal{V}(x)$ in the fermionic case in (j-l) are only slightly affected by a different ramping rate. The input files of the simulations are given as simulations \#20 and \#21 in Table~S4 of the supplementary material~\cite{SUPPLMAT_TUTORIAL}.Video showing the evolution of all these one-body correlation functions in time is available on \url{https://www.youtube.com/watch?v=l2UsTPmJ6po}.   } \label{fig:dyn_bosons_corr}
\end{figure}

Since the system follows the ground state in the slow limit (large $\tau$) with an even number of particles, we now investigate the effect of a quench (smaller $\tau$) using $N=6$ bosons or fermions. We ramp up the barrier from zero to $E_\text{dw}=20$ in three different ramping times, $\tau=1$, $\tau=10$ and $\tau=50$, to which we refer as a fast, an intermediate, and a slow ramp, respectively, in the following. Their one-body correlation functions at a given time, $t=80$, are shown in Fig.~\ref{fig:dyn_bosons_corr}(a-c) for bosons and in Fig.~\ref{fig:dyn_bosons_corr}(g-i) for fermions.  

With a slow ramp ($\tau=50$), the ground state of the final Hamiltonian is recovered -- the correlation between the two wells vanishes [cf. areas where $x$ is to the left (right) of the barrier and $x'$ to its right (left) and $\vert g^{(1)}(x,x')\vert \approx 0$ in Fig.~\ref{fig:dyn_bosons_corr}(c,i)].
As the ramping time becomes shorter $\tau=10$, fluctuations appear on the correlation function, indicating excitations [cf. areas where $x$ is to the left (right) of the barrier and $x'$ to its right (left) and $\vert g^{(1)}(x,x')\vert > 0$ in Fig.~\ref{fig:dyn_bosons_corr}(b,h)].
These excitations accumulate rapidly as the ramping time shortens and, eventually, with a fast ramping $\tau=1$, we obtain a strongly fluctuating one-body correlation function [Fig.~\ref{fig:dyn_bosons_corr}(a,g)].

To analyze the quantum fluctuations in the time-evolving states are affected by the speed of the ramp, we quantify the density fluctuations using the position-dependent variance $\mathcal{V}(x)$ [Eq.~\eqref{eq:var}] extracted from $10,000$ single-shot simulations for bosons in Fig.~\ref{fig:dyn_bosons_corr}(d),(e),(f) and for fermions in Fig.~\ref{fig:dyn_bosons_corr}(j),(k),(l). 

Generally, we find the fluctuations are large where the density is large. For bosons, a slower ramp results in an increase of the magnitude of fluctuations [compare Fig.~\ref{fig:dyn_bosons_corr}(d),(e),(f)]. For fermions, the magnitude of fluctuations is practically unaffected by the pace of the ramp [compare Fig.~\ref{fig:dyn_bosons_corr}(j),(k),(l)].
We infer that the behavior of the magnitude of the fluctuations is analogous to the correlation order parameter (COP) $\Delta$ [Eq.~\eqref{eq:fragmentation_OP}]: for bosons, the COP increases a lot as the ramping rate decreases in long time $t$ [cf. Fig.~\ref{fig:bosons_hysteresis}(g)], whereas the COP varies only very slightly at different ramping rates for the case of fermions [cf. Fig.~\ref{fig:bosons_hysteresis}(h)].
This analogous behavior of the quantum fluctuations and the correlation order parameter has previously been used to extract the phases of dipolar ultracold atoms in lattices~\cite{chatterjee:18,chatterjee2:19}. 

The excitations generated by the quench thus also manifest themselves in the eigenvalues of the reduced one-body density matrix or orbital occupations as represented by the COP  and its time-dependence [Fig.~\ref{fig:bosons_hysteresis}(g,h)]. 
For bosons, in the case of a slow ramp, only two orbitals are macroscopically occupied as in the ground state, while in the case of a fast ramp, the contribution of the first $10$ orbitals are in the same order of magnitude and the system is thus highly correlated. The extra fragmentation comes from the dynamics of the system. 

For bosons, in the slow case, the COP decreases as the barrier increases and converges to $\Delta=0.5$ as in the ground state [cf. Fig.~\ref{fig:ground_occupation_OP}(a)]. As the ramping becomes faster, the COP starts to oscillate and the amplitude becomes larger and larger. However, the dynamical behavior of COP becomes qualitatively different as the ramping time becomes shorter than $\tau=10$. It no longer oscillates in a regular manner but rather decreases rapidly and fluctuates. This implies that many high-energy eigenstates of the final Hamiltonian are populated in the dynamics of the system.

For fermions, the absolute values of the orbital occupations are not drastically influenced by the system dynamics and neither is the COP. 
However, the dynamical behavior of the COP is completely different in the fast and slow ramping limits -- 
in contrast to the bosonic case, the COP increases slightly for a slow ramp, moving closer to the value for non-interacting fermions $\Delta=0.167$. This indicates that the effects of the interparticle interactions become weaker since the fermions are now divided into two well-separated groups.

The results for the one-body correlation, the COP, and the single-shot simulations point towards a dynamical phase transition at roughly $\tau_C\sim10$ for fermions and bosons alike. The system behaves like the ground state for ramps slower than $\tau_C$, but becomes highly excited for ramps faster than $\tau_C$.

\subsection{Conclusion and Discussions}\label{sec:conc}

In this work, we have described and demonstrated the workflow, usage and structure of the MCTDH-X software hosted, documented and distributed through \url{http://ultracold.org}. In a step-by-step tutorial, we show how to examine particle correlations and fluctuations using the MCTDH-X software. We exhibited applications of MCTDH-X to solutions of the time-dependent and time-independent many-body Schr\"{o}dinger equation for systems of bosons and fermions in a double-well potential. We have computed the many-body wavefunction and calculated various quantities of interest, including correlations, real- and momentum-space density distributions and single-shot experiments. We highlight that our paper demonstrates the first application of MCTDH-X to obtain single shot images for fermions, to analyze correlations via single shot counting statistics, to investigate the dynamical behavior of the correlation order parameter, and to directly compare the dynamical behavior of bosons and fermions using single-shot images and the correlation order parameter.

The double-well potential captures many salient features of the Hubbard model, like the superfluid-Mott insulator transition of bosons. Additionally, we find a crystal state emerging for fermions that bears some similarities with strongly interacting bosons that are en route to the fermionization limit.

In the ground state, we thus have already observed a wide range of states of fermions and bosons by tuning the interparticle interactions and the barrier height of the double well. Apart from the transition into the Mott phase, the bosonic system approaches the non-interacting fermionic one as the interparticle interactions increase towards the fermionization limit. Fermionized bosonic states have many similarities to non-interacting fermions in the real-space representation.
We demonstrate that the superfluid, Mott insulating, fermionized and crystallized phases of many-body states can be distinguished by their correlation functions and by a correlation order parameter that is a function of the eigenvalues of the reduced one-body density matrix.

Dynamics of a system may drive it out of the ground state and induce quantum fluctuations and correlations. 
We use MCTDH-X to solve the time-dependent Schr\"{o}dinger equation for bosons and fermions in a one-body potential that is smoothly transformed from a single into a double well by ramping up a Gaussian barrier at different time scales. 
We find a novel and unexpected hysteretic behavior that heralds a first-order phase transition for the case of an odd number of bosons or fermions. When the particle number is even, there is no hysteresis, heralding a second-order phase transition. We thus demonstrate that the order of the superfluid-Mott insulator transition depends on the number of particles in our finite-size double-well system. 
With an odd number of particles, strong residual correlations between the two wells prevail even in the limit of large barriers for any pace of the ramp of the barrier. This implies that the superfluidity of the system cannot be eliminated by increasing the barrier and the system cannot enter the Mott insulator phase dynamically. 

For faster ramps, our tutorial example approaches a quench scenario: when the double well's barrier is quenched up rapidly, a considerable amount of eigenstates of the final Hamiltonian become relevantly occupied and participate in the emergent quantum dynamics. This suggests strong fluctuations, which can be confirmed by an analysis of the variance in single-shot images. The significant differences between the fast and slow ramps that we observe point to a dynamical phase transition.

Finally, we note that although we only show examples with a few particles in this tutorial, the MCTDH-X is able to provide many-body states of systems with a much larger number of particles~\cite{klaiman:15,alon:19}. 

\acknowledgements{We acknowledge the financial support from the Swiss
	National Science Foundation (SNSF), the ETH Grants,
	Mr. Giulio Anderheggen, the Austrian Science Foundation (FWF) under grants P32033 and M2653. We also acknowledge
	the computation time on the ETH Euler and the HLRS Hazel Hen clusters.}

\bibliography{MCTDHX_Mendeley}

\end{document}


\thispagestyle{empty}
	\begin{center}
		
		\includegraphics[width=0.6\textwidth]{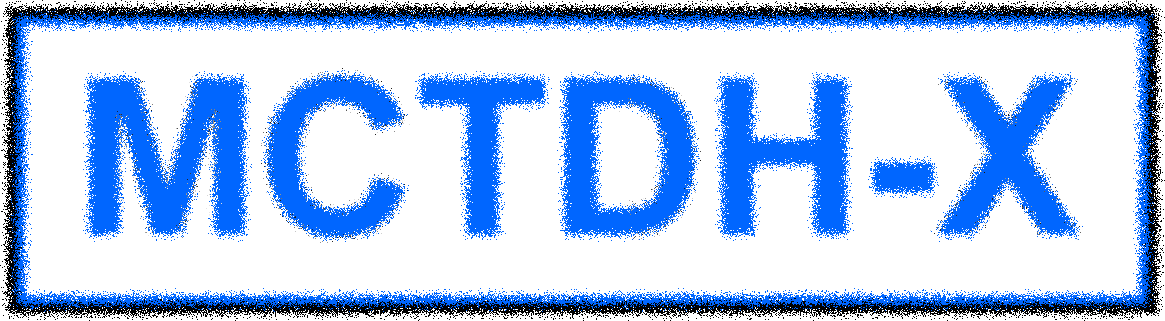}\\
		\Large{Supplementary Material: \textit{MCTDH-X}: The multiconfigurational time-dependent Hartree method for indistinguishable particles software
			}\\[2cm]
		\Large{{\bf \htmladdnormallink{https://arxiv.org/abs/1911.00525}{https://arxiv.org/abs/1911.00525}}}\\[5mm]
			\normalsize{by Rui Lin, Paolo Molignini, Luca Papariello, Marios C. Tsatsos, Camille L\'ev\^eque, Storm E. Weiner, Elke Fasshauer, R. Chitra, and Axel U.~J. Lode}\\[1cm]
		\Large{{\bf This document: \htmladdnormallink{http://ultracold.org/tutorial}{http://ultracold.org/tutorial}}}
		\\[2cm]
		\Large{{\bf Contact: \htmladdnormallink{mctdhx@ultracold.org}{mailto:mctdhx@ultracold.org}}\\[5mm] 
			{\bf Download and Support: \htmladdnormallink{http://ultracold.org/forum}{http://ultracold.org/forum}}}
	\end{center}

	\clearpage

In this supplemental material, we give details on how to obtain the results and visualizations presented in the main text in Sec.~\ref{sec:redo}. This guide to reproduce the results of our work is accompanied by video tutorials at ~\htmladdnormallink{https://www.youtube.com/playlist?list=PLJIFUqmSeGBKxmLcCuk6dpILnni\_uIFGu}{https://www.youtube.com/playlist?list=PLJIFUqmSeGBKxmLcCuk6dpILnni\_uIFGu}. 
A rough guide to doing customized simulations is provided in Sec.~\ref{sec:custom}.
A discussion on the convergence of MCTDH-X predictions is given in Sec.~\ref{sec:orbital_converge}, and the physics of the Bose-Hubbard model and its comparison to MCTDH-X are discussed in Sec.~\ref{sec:BH}. This supplemental material is hosted as a webpage at \htmladdnormallink{http://ultracold.org/tutorial}{http://ultracold.org/tutorial}.

\section{Doing the MCTDH-X simulations in the main text}\label{sec:redo}

\subsection{Download and installation of the MCTDH-X software}
\subsubsection{Download}
We restrict our exposition of the download and installation to a minimum here. A detailed description is given in the MCTDH-X manual \htmladdnormallink{http://ultracold.org/data/manual}{http://ultracold.org/data/manual}. 
The software can be downloaded from the repository using the command: 

\texttt{hg clone https://bitbucket.org/MCTDH-X/mctdh-x-release}

The username and password for the repository can be obtained at \htmladdnormallink{http://ultracold.org/forum}{http://ultracold.org/forum} .

\subsubsection{Compilation and installation}

For the prerequisites of the installation of the MCTDH-X software, please look at the manual
at\linebreak \htmladdnormallink{http://ultracold.org/data/manual}{http://ultracold.org/data/manual}.

Navigate to the folder \texttt{mctdh-x-release} and run \texttt{./Install\_MCTDHX.sh}.
Follow the installation and answer the questions according to your computer / cluster setup.
If the installation is successful, there will be the executables \texttt{MCTDHX\_analysis\_<compiler>}, \texttt{MCTDHX\_<compiler>}, in the folder \texttt{mctdh-x-release/bin}
(here \texttt{<compiler>} will be either \texttt{intel} or \texttt{gcc}.

\subsection{Repeating the simulations in the main text}

Below, we discuss three alternative ways to redo the computations in this tutorial in Secs.~\ref{sec:scripts}, \ref{sec:inputs}, 
and \ref{sec:scratch}.

Sec.~\ref{sec:scripts} describes how to redo all computations using supplied bash scripts and provides the easiest 
access, however, with little ``under-the-hood'' knowledge of the MCTDH-X software. Sec.~\ref{sec:scripts} provides
a directory tree with scripts that perform the computations and their visualization; this is a good starting point for
exploring what happens, if parameters in the presupplied inputs are modified, see below and \htmladdnormallink{http://ultracold.org/manual}{http://ultracold.org/manual} for help.

Sec.~\ref{sec:inputs} describes how to redo the tutorial computations using the input files collected in this 
archive. Following our tutorial provides an understanding of what files are necessary to run a (series 
of) computation(s) with the MCTDH-X software.

Sec.~\ref{sec:scratch} is a rough guide on how to redo the computations in the tutorial ``from scratch'': the inputs
are obtained by modifying the default input files distributed with the MCTDH-X software -- the scripts 
and inputs in this archive are not used at all (or for reference purposes only). Eventually, doing 
(part of) the tutorial examples in this way provides the qualification to an independent use of the 
MCTDH-X software.

\subsubsection{Using the provided scripts}\label{sec:scripts}

The included bash scripts are named in analogy to the figures in the manuscript to which they correspond, 
i.e., \texttt{FIG2\_3.sh}, \texttt{FIG4.sh}, \texttt{FIG5.sh}, \texttt{FIG6.sh}. See folder tree structure in Figs.~\ref{fig:tree1}, \ref{fig:tree2}, \ref{fig:tree3_1}, \ref{fig:tree3_2}, \ref{fig:tree3_3}, \ref{fig:tree4_1} and \ref{fig:tree4_2} below.

They can be executed as follows:
\begin{enumerate}
\item \texttt{./FIG2\_3.sh <X>}
\item \texttt{./FIG4.sh <X> }
\item \texttt{./FIG5.sh <X> }
\item \texttt{./FIG6.sh <X> }
\end{enumerate}

Here, \texttt{<X>} represents the number of processes to use for performing the computation. Usually, replacing
\texttt{<X>} by $4$ is a good choice, if your system has more than four CPU cores.
The above scripts will also automatically create the visualizations of the data corresponding to Figs.~2 to 6. in the manuscript.

\subsubsection{Using the provided input files}\label{sec:inputs}

All inputs necessary to repeat the computations in the manuscript are included (see tree structure in Figs.~\ref{fig:tree1}, \ref{fig:tree2}, \ref{fig:tree3_1}, \ref{fig:tree3_2}, \ref{fig:tree3_3}, \ref{fig:tree4_1} and \ref{fig:tree4_2}  below).

To repeat the computations in the manuscript using the provided inputs, the following steps can guide you:

\begin{enumerate}
\item Select a figure and input from the manuscript and navigate to the corresponding directory. 
For instance, for Fig.2 (a)--(c):
\texttt{cd Relax1\_FIG2abc}

\item \emph{Optionally:} If you restart your computation from a previous one in the folder \texttt{<restart-from>}, copy the necessary 
data from there to the working directory (this applies to examples Fig.~5(e), \#13--\#17 , \#20 and \#21 where \texttt{GUESS=\textquotesingle BINR\textquotesingle} in Tables~\ref{table:input_file_prop} and \ref{table:input_file_prop2}), and:\\
\texttt{cp <restart-from>/*bin}\\
\texttt{cp <restart-from>/Header}

\item Copy the executables of the MCTDH-X software: \texttt{bincp}

\item Run the program: \texttt{MCTDHX}

\item Run the analysis if it is needed: \texttt{MCTDHX\_analysis}

\item Plot the results with gnuplot, python or other visualization tools.\\
\end{enumerate}

In order to re-generate the visualizations shown in the manuscript, the python scripts \texttt{FIG2.py}, ..., \texttt{FIG6.py} can be adapted such that they reflect the folder tree created. Alternatively, custom visualizations can be done by resorting to the output description in the manual
at \htmladdnormallink{http://ultracold.org/manual}{http://ultracold.org/manual}.

\subsubsection{Doing the tutorial examples ``from scratch''}\label{sec:scratch}

To prepare the input files in the steps 4. and 6. below, please confer the Tables~\ref{table:input_file_relax}, \ref{table:input_file_series}, \ref{table:input_file_prop}, \ref{table:input_file_prop2} and \ref{table:input_file_analysis}. These tables illustrate how to modify the default input files included in the MCTDH-X software in order to repeat the simulations that are presented in the main text. For a detailed description of the inputs, refer to the manual at \htmladdnormallink{http://ultracold.org/manual}{http://ultracold.org/manual}.

To run the MCTDH-X software, the following steps are generally taken:

\begin{enumerate}
    \item Create a new folder. 

	\item \emph{Optionally:} If you restart your computation from a previous one in the folder \texttt{<restart-from>}, copy the necessary 
	data from there to the working directory:\\
	\texttt{cp <restart-from>/*bin}\\
	\texttt{cp <restart-from>/Header}

	\item \emph{Optionally:} If you do not restart a previous computation copy the default inputs and executables:\\
	\texttt{inpcp}\\
	\texttt{bincp}

    \item Modify the input file \texttt{MCTDHX.inp} according to your needs (for instance, as indicated in the Tables~\ref{table:input_file_relax},\ref{table:input_file_series}, \ref{table:input_file_prop}, \ref{table:input_file_prop2} below).

	\item Run the program: \texttt{MCTDHX}

    \item Configure the analysis task by modifying the file \texttt{analysis.inp} to extract the observables you want (for instance, as indicated in Table \ref{table:input_file_analysis} below).	

	\item Run the analysis: \texttt{MCTDHX\_analysis}
	
	\item Visualize the results using \texttt{gnuplot} or similar. For a detailed description on the structure of the input and output of the MCTDH-X software, please refer to the manual at \htmladdnormallink{http://ultracold.org/manual}{http://ultracold.org/manual}.
\end{enumerate}

For the relaxations, the program (command \texttt{MCTDHX} ) can be directly executed when the executables and the modified input file that contains the Hamiltonian are present in the current working directory. For relaxations, we use randomly generated initial states to circumvent a convergence to excited states. When ``repetition'' is given as a keyword in a table, the program should be executed multiple ($\sim$10) times with the same input file, and the energies of the resulting states should be compared in order to choose the correct ground state (this applies to problem \#10, i.e., Fig. 5(a) and Table ~\ref{table:input_file_prop}). 

In our simulations of time-evolving states, the initial states are all obtained from relaxations. The results of the relaxations are saved in the binary files \texttt{PSI\_bin}, \texttt{CIc\_bin} and \texttt{Header}. They should be copied into the folder where propagations are performed. With the input file, binary files, and library in one folder, we can execute the program with the command \texttt{MCTDHX}. We should always verify that the propagations and the corresponding relaxations have compatible input variables like orbital number, particle number, grid size, etc.. 

There are several input variables for which we provide a more detailed explanation.

\begin{itemize}
	\item The potentials \texttt{HO1D}, \texttt{HO1D+td\_gauss}, and \texttt{HO1D+td\_g\_r} are defined respectively as
	\begin{subequations}
		\begin{eqnarray}
		V(x) &=& \frac{1}{2}p_1^2(x-p_2)^2,\quad  \mathtt{HO1D} \\
		V(x) &=& \begin{cases}
		\dfrac{1}{2}p_1^2(x-p_2)^2 + p_3\dfrac{t}{p_4}\exp\left[ -(x-p_5)^2/2p_6^2 \right], t<p_4 \\
		\dfrac{1}{2}p_1^2(x-p_2)^2 + p_3\exp\left[  -(x-p_5)^2/2p_6^2 \right], t\ge p_4
		\end{cases},\quad  \mathtt{HO1D+td\_gauss} \\
		V(x) &=& \begin{cases}
		\dfrac{1}{2}p_1^2(x-p_2)^2 + p_3\left(1-\dfrac{t}{p_4}\right)\exp\left[ -(x-p_5)^2/2p_6^2  \right], t<p_4 \\
		0, t\ge p_4
		\end{cases},\quad  \mathtt{HO1D+td\_g\_r} 
		\end{eqnarray}
	\end{subequations}
	where $x$ is the position, $t$ is the imaginary or real time, and $p_1$ to $p_6$ correspond to the input variables \texttt{parameter1} to \texttt{parameter6} in the input file \texttt{MCTDHX.inp}, respectively. 
	
	\item The interparticle interaction can be chosen in the following way. The contact interaction can be chosen by setting \texttt{Interaction\_Type} = 0. In this case, the variables \texttt{which\_interaction}, \texttt{Interaction\_Parameter1}, and \texttt{Interaction\_Parameter2} are irrelevant. The regularized Coulomb interaction can be chosen by setting \texttt{which\_interaction} = \texttt{\textquotesingle regC\textquotesingle}, \texttt{Interaction\_Type} = 4, and \texttt{Interaction\_Parameter1} and \texttt{Interaction\_Parameter2} are respectively $\alpha$ and $\beta$ in Eq.(14b) in the main text. Other type of interaction can be chosen by using different keywords for \texttt{which\_interaction}. Detailed descriptions of these interactions are given in the manual at \htmladdnormallink{http://ultracold.org/data/manual}.
	\item The number of grids \texttt{NDVR\_X} is recommended to be a powers of 2.
\end{itemize}

If you get stuck, you may check the  supplied input files and scripts in Sec.~\ref{sec:scripts} available for download at \htmladdnormallink{http://ultracold.org/data/tutorial\_input\_files.zip}{http://ultracold.org/data/tutorial_input_files.zip} and the manual at \htmladdnormallink{http://ultracold.org/manual}{http://ultracold.org/manual}. 
The structure of the folder tree in \texttt{tutorial\_input\_files.zip} is explained in Figs.~\ref{fig:tree1},\ref{fig:tree2},\ref{fig:tree3_1}, \ref{fig:tree3_2}, \ref{fig:tree3_3}, \ref{fig:tree4_1}and \ref{fig:tree4_2}.

\subsubsection{Results}

The simulation results are saved in the binary files \texttt{PSI\_bin}, \texttt{CIc\_bin} and \texttt{Header}.
In certain time intervals set by the input variable \texttt{Output\_TimeStep} in \texttt{MCTDHX.inp}, the instantaneous 
results of a few important quantities of interest are created by running \texttt{MCTDHX}.
In the files \texttt{*.*000orbs.dat} (these files are output by calling \texttt{MCTDHX}, if the \texttt{Write\_ASCII} parameter 
in \texttt{MCTDHX.inp} is set to be \texttt{.T.}), the first three columns are the positions in three dimensions. 
The fifth column is the instantaneous one-body potential $V(x;t)$. The sixth column is the instantaneous 
density distribution. See the manual at \htmladdnormallink{http://ultracold.org/manual}{http://ultracold.org/manual} for a detailed description of the output.

The file \texttt{NO\_PR.out} contains the orbital occupations (eigenvalues of the reduced density matrix) and 
system energy.  For relaxations, the values in the \texttt{NO\_PR.out} should converge for a reasonable and 
meaningful result.

Further quantities of interest are extracted using the command \texttt{MCTDHX\_analysis} configured via \texttt{analysis.inp} (see
table \ref{table:input_file_analysis} below).

For further reference, please use the detailed description of the input and output files in the MCTDHX manual at 
 \htmladdnormallink{http://ultracold.org/data/manual}{http://ultracold.org/data/manual}

\subsection{Folder structure}

The folders are named referencing both
\begin{enumerate}
\item the number label of simulations in the Tables~\ref{table:input_file_relax}, \ref{table:input_file_series}, \ref{table:input_file_prop}, \ref{table:input_file_prop2} below, and 
\item the figures for which the folders are relevant in the main text. Relaxations and propagations are 
grouped by relevance to certain figures as specified in the folder tree shown in Figs.~\ref{fig:tree1},\ref{fig:tree2},\ref{fig:tree3_1}, \ref{fig:tree3_2}, \ref{fig:tree3_3}, \ref{fig:tree4_1}and \ref{fig:tree4_2}  below. 
\end{enumerate}
Relevant relaxations and propagations are grouped in the same folder. 
For example, simultaion \#9 is a relaxation which prepares the initial state for the propagation \#15, 
whose result (off-diagonal of one-body correlation function) is shown in Fig.5(f) in the main text. 
The input files of problems \#9 and \#15 are thus given in the folder \texttt{Relax9\_Prop15\_FIG5f} (see below Fig.~\ref{fig:tree3_1}).

\subsubsection{Folder contents}

In each (sub-)folder, the input file for an MCTDH-X computation \texttt{MCTDHX.inp}, i.e., a relaxation or propagation and (optionally) an analysis input file \texttt{analysis.inp} are given. 

In particular:
\begin{enumerate}
\item In the subfolders \texttt{prop\_a}, \texttt{prop\_b}, \texttt{prop\_c} of \texttt{Relax18\_Prop20\_FIG6a-f}, and subfolders \texttt{prop\_g}, \texttt{prop\_h}, \texttt{prop\_i} of \texttt{Relax19\_Prop21\_FIG6g-l/prop\_g}, we provide two analysis input files. 
The first one, \texttt{analysis.inp}, generates correlation functions and single-shot images at $t = 80.0$ for Fig.~6 in the main text. 
The second one, \texttt{analysis2.inp}, generates correlation functions from $t = 0$ to 
$t = 80$ for the video provided at~\htmladdnormallink{https://www.youtube.com/watch?v=l2UsTPmJ6po}{https://www.youtube.com/watch?v=l2UsTPmJ6po}. These input files need to be renamed to \texttt{analysis.inp} when they are used via \texttt{MCTDHX\_analysis}. This procedure is automatized by the bash script \texttt{movie.sh}.
 (see Fig.~\ref{fig:tree4_1}, \ref{fig:tree4_2} and \ref{fig:tree5}).

\item In the folder \texttt{Relax10\_Prop16\_FIG5ae/relax}, due to the presence of low-lying excited states, the relaxation needs to be run multiple times to ensure that the ground state has been obtained. Due to the random initial conditions, the different simulations generate different results. We need to compare the energies of the results to choose the true ground state. This procedure is automatized by the shell script \texttt{Run\_multiple.sh} (cf. Fig.~\ref{fig:tree3_2} and label ``repetition'' in Table \ref{table:input_file_prop}).

\item In the folders \texttt{Relax\_series\_FIG5f\_FIG4a}, \texttt{Relax\_series\_FIG4b}, 
	\texttt{Relax\_series\_FIG4c} and \texttt{Relax\_series\_FIG5e}, we provide a bash script \texttt{parameter\_sweep.sh} for running parameter sweep for 
	the variable \texttt{parameter3} ($E_\text{dw}$) or \texttt{xlambda\_0} ($g$).

\item In the folder \texttt{Relax\_series\_FIG5e}, due to the presence of low-lying energy states, we need to simulate multiple times and choose the ground state at each parameter point. To save time, we use the following method instead. We first choose a certain parameter point, e.g. \texttt{parameter3} = 20.d0, and simulate multiple times. We will obtain two completely different kinds of states: one of them has vanishing correlation between the two wells and the other one has finite correlation between wells. These two states are provided in the subfolders \texttt{relax\_series1} and \texttt{relax\_series2}, respectively, as binary files \texttt{PSI\_bin}, \texttt{CIc\_bin}, and \texttt{Header}.

The bash scripts \texttt{parameter\_sweep.sh} in the two subfolders then use these two states as initial states and do relaxations at different barrier heights $E_\mathrm{dw}$. The basic characteristics, i.e., whether the correlations between the two wells are vanishing or not, remains unchanged during relaxations.
As the barrier height changes, these two kinds of states compete to be the simulation result. At each parameter point, we compare the results in \texttt{relax\_series1} and \texttt{relax\_series2}, and choose the correct ground state.

\end{enumerate}

\section{Creating your own MCTDH-X simulations}\label{sec:custom}

After redoing the simulations provided in this tutorial, the readers are encouraged to create their own simulations tailored to their own specific problems at hand. In this case, the following steps should be taken:

\begin{enumerate}
	\item Create a working folder, copy the exemplary input file \texttt{MCTDHX.inp} into the working folder using command \texttt{inpcp} and modify it according to the following steps:

	\begin{enumerate}
	\item Determine the type of the problem, i.e., whether it is static (ground state) or dynamic (time-evolving state).  Related input variables in \texttt{MCTDHX.inp}: \texttt{Job\_Prefactor}.
	
	\item Determine the Hamiltonian of the problem, i.e., determine the number and statistics of the particles, the one-body potential and the two-body interaction. Related input variables in \texttt{MCTDHX.inp}: \texttt{JOB\_TYPE}, \texttt{Npar}, \texttt{whichpot}, \texttt{parameter1}, \texttt{parameter2}, $\ldots$, \texttt{xlambda\_0}, \texttt{which\_interaction}, \texttt{Interaction\_Type}, \texttt{Interaction\_Parameter1}, \texttt{Interaction\_Parameter2} ... Customized potentials and interactions can be created using the source files \texttt{Get\_1bodyPotential.F} and \texttt{Get\_InterParticle\_Potential.F}, respectively.
	
	\item Determine the other physical parameters used in the simulation, including the number of orbitals, the dimension of the system, the spatial extent, the spatial grid, etc. Related input variables in \texttt{MCTDHX.inp}: \texttt{Morb}, \texttt{DIM\_MCTDH}, \texttt{NDVR\_X}, \texttt{x\_initial}, \texttt{x\_final}, etc.
	
	\item Determine the initial state of the problem. This is usually done by using a randomly generated initial state or preparing the initial state with another simulation. In the latter case, the user should 
	set up the ground state or a time-evolving state of a particular Hamiltonian as initial state	
	with another simulation directory. Then the user should copy the generated files \texttt{Header}, \texttt{PSI\_bin} and \texttt{CIc\_bin} into the working folder. Related input variables in \texttt{MCTDHX.inp}: \texttt{GUESS}, \texttt{Binary\_Start\_Time}.
	
	\item Determine the integration parameters. Related input variables in \texttt{MCTDHX.inp}: \texttt{Time\_Begin}, \texttt{Time\_Final}, \texttt{Output\_TimeStep}, \texttt{Integration\_Stepsize}, \texttt{Coefficients\_Integrator}.

	\end{enumerate}	

	\item Copy the executables \texttt{MCTDHX\_gcc}/\texttt{MCTDHX\_intel} and \texttt{analysis\_gcc}/\texttt{analysis\_intel} into the working folder using command \texttt{bincp}. Run the executable using command \texttt{MCTDHX}.
	
	\item Sanity test of the result

	\begin{enumerate}
	\item Check the one-body potential in \texttt{***orbs.dat} to see if they are correctly generated as specified by the Hamiltonian. Check the evolution of the energy and orbital occupations as functions of real/imaginary time in the generated file \texttt{NO\_PR.out} and the evolution of the real-space density distribution in \texttt{***orbs.dat} to see if they are physically sensible. If not, confirm that there is no mistake/typo in the input file.
	
	\item Check if the real-space density distribution vanishes at the boundary. If not, use a wider spatial extent.
	
	\item Test the convergence in different parameters. Use a different spatial grid and orbital number and check if the results remain unchanged. A detailed discussion on the convergence in orbital number can be found in Sec.~\ref{sec:orbital_converge} and in Ref.~\cite{lode:12,lode1:16}.
	\end{enumerate}	

	\item Analysis of the result
	\begin{enumerate}
		\item Specify the desired time interval of analysis. Related input variables in \texttt{analysis.inp}: \texttt{Time\_From}, \texttt{Time\_To}.
		
		\item Specify the desired quantities of interest. Refer to the manual at \htmladdnormallink{http://ultracold.org/data/manual}{http://ultracold.org/data/manual} for related input variables in \texttt{analysis.inp}.
		
		\item Visualize the generated quantities of interest.
	\end{enumerate}
\end{enumerate}


\section{Discussion on the convergence in the orbital number}\label{sec:orbital_converge}

The convergence in the orbital number is an important issue in MCTDH-X. The choice of the orbital number is a trade-off between computation speed and simulation accuracy. Without enough orbitals, we will not only obtain incorrect particle correlations, but sometimes even an incorrect density distribution. 
The simplest method to determine the convergence is by comparing the occupations. A higher-order orbital is considered insignificant, when its occupation is significantly lower than the lower-order orbitals. This criterion sometimes fails when the seemingly insignificant orbitals contain important information about particle correlations and fluctuations of the simulated many-body state. 

An intuitive example is strongly-interacting bosons. Fig.~\ref{fig:corr2_diff_M}(a) shows the  real-space density distribution of $N=2$ bosons with extremely strong interaction $g=500$ simulated with different orbital number. When $M=1$ orbital is used, the interaction can only be captured in the Gross-Pitaevskii mean-field manner, and the many-body state simply reduces to a Thomas-Fermi cloud. As soon as the second orbital is used, the double-peak density distribution can be captured, though an obvious discrepancy with the fermionic distribution can be observed. Such a discrepancy diminishes as more orbitals are used. 

A more intuitive discussion on the convergence of MCTDH-X results with the orbital number can be illustrated by the comparison between the standard single-band Bose-Hubbard model and the simulation results for bosons in a Mott insulator state.

\section{Bose-Hubbard model and beyond with MCTDH-X}\label{sec:BH}

The Hubbard model usually serves as good approximation for a lattice system. It assumes only one Wannier function contributes in each lattice site. 
We note that the MCTDH-X enables a modelling of the many-body state with an improved accuracy as compared to the Hubbard-lattice-models, because it uses a general, variationally optimized and not necessarily site-local basis set. This improved accuracy is of particular importance in cases where the single-band Hubbard description fails, for instance, due to the action of a blue-detuned cavity~\cite{lin2:19} or long-range dipolar interactions~\cite{chatterjee:18,chatterjee:19,chatterjee2:19}. Direct comparisons between the MCTDH-X and Hubbard models have been performed in Refs.~\cite{dutta:19,lode:15,mistakidis:14,mistakidis:15,neuhaus-steinmetz:17,mistakidis:17,sakmann:09,sakmann:11,sakmann:10,lin:19}.

The Bose-Hubbard Hamiltonian takes the following form in one dimension,
\begin{eqnarray}
\hat{H}_\text{BH} = -t\sum_{\langle i,j\rangle} (\hat{c}^\dagger_i \hat{c}_j+\text{H.c.})+ U\sum_i \hat{n}_i(\hat{n}_i-1),
\end{eqnarray}
where, $\hat{c}_i$ is the annihilation operator for a particle in the Wannier state $w_i(x)$ that is localized at site $i$ of the lattice and $\hat{n}_i=\hat{c}^\dagger_i\hat{c}_i$ is the corresponding number operator. In the Hubbard model, the superfluid to Mott-insulator transition is a result of the competition between the hopping strength $t\propto \int dx w^*_i(x) \partial_x^2w_{i+1}(x+a)$ between lattice sites with lattice constant $a$, and the on-site interaction $U\propto\int dx |w_i(x)|^4$ between particles. We consider a system with a fixed total number of particles $N$ and therefore neglect the chemical potential.

In the limit $t/U\rightarrow\infty$, the system is in the superfluid phase, and the ground state is given by $|\Psi\rangle \propto \left(\sum_i a^\dagger_i\right)^N|0\rangle$. In this state, the $U(1)$ symmetry $\hat{c}_i\rightarrow e^{i\theta}\hat{c}_i$ of the system is broken spontaneously, such that all the particles have the same phase. Particles in different sites are coordinated; they are \textit{condensed} in the same single-particle state. This condensation entails a Glauber one-body correlation between sites which is unity.
In the limit $U/t\rightarrow\infty$, the system is in a Mott insulator phase where the ground state of the system is given by $|\Psi\rangle \propto \prod_i \left(a^\dagger_i\right)^{n_i}|0\rangle$, where $n_i$ is the number of particles in the $i$-th site and satisfies the constraint $\sum_in_i=N$. In this state, particles in different sites are disconnected and have no information on each other's behaviors.  The phases of the particles in different sites lose coherence. Strong correlation effects build up and the one-body correlation between sites vanishes. 
The transition point given in units of $t/U$ depends on the structure of the lattice. In particular, it is roughly proportional to the filling factor and the inverse of the dimensionality~\cite{teichmann:09}.

\begin{figure}[t]
	\centering
	\includegraphics[width=\textwidth]{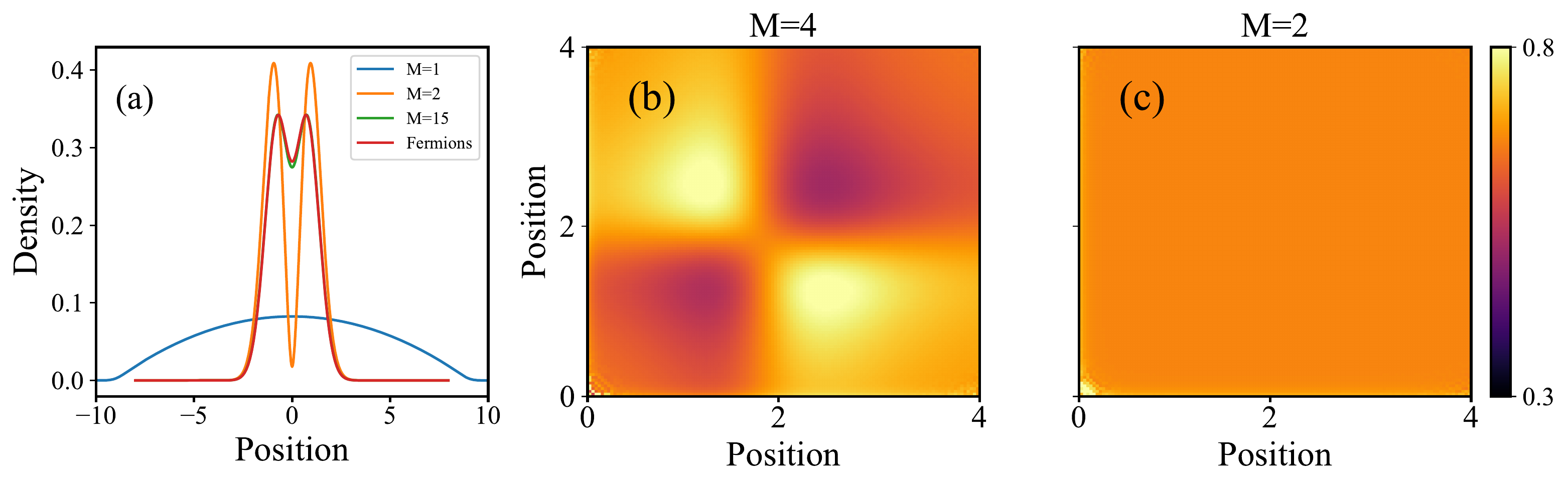}
	\caption{(a) The real-space density distribution for $N=2$ bosons with extremely strong interactions $g=500$ performed with different orbital numbers $M=$1, 2, 15. As the number of orbitals increases, the simulated density profile becomes closer and closer to the density profile of fermions.
	(b,c) The two-body correlation function inside one well $g^{(2)}(x>0,x'>0)$ of a double-well potential simulated with $N=6$ bosons and (b) $M=6$ orbitals (c) $M=2$ orbitals. More details are unveiled by the extra orbitals in panel (a), indicating physics beyond the single-band Hubbard model.} \label{fig:corr2_diff_M}
\end{figure}

For illustration, we consider $N=6$ weakly-interacting ($g=1$) bosons in a double-well potential with high barrier $E_\text{dw}=20$, where the two wells are Mott insulating from each other. 
According to the single-band Bose-Hubbard model, the energy of this system is minimzed when the bosons are evenly distributed into the two wells, with three bosons in each well. The many-body state can thus be written as
$|\Psi\rangle \propto \left(\hat{c}^\dagger_L\right)^3\left(\hat{c}_R^\dagger\right)^3|0\rangle$,
where $\hat{c}^\dagger_L$ and $\hat{c}^\dagger_R$ are the creation operators for the band in the left and right well, respectively. Although such a picture is well supported by the momentum space density distribution and  the one-body correlation function [$g^{(1)}$][cf. Fig.~2(e,f) in the main text], it contradicts with the two-body correlation function [$g^{(2)}$]  obtained from the simulations. In a Mott insulator state where $g^{(1)}$ between lattice sites vanishes, the diagonal term of $g^{(2)}$ is directly related to the number of particles $n_i$ in the corresponding site through $g^{(2)}(x,x) = 1-1/n_i$~\cite{lin:19}. If the single-band Bose-Hubbard model holds true, the diagonal term of $g^{(2)}$ would be uniformly $g^{(2)}(x,x) = 2/3$ since $n_i=3$. This is not the case in the simulations when sufficiently many orbitals are used. The landscape of the correlation function is uneven, with values ranging from 0.4 to 0.8 [Fig.~\ref{fig:corr2_diff_M}(b)].  On the contrary, if only $M=2$ orbitals are used in the simulations, indeed $g^{(2)}(x,x) = 2/3$ holds uniformly [Fig.~\ref{fig:corr2_diff_M}(c)]. These results show that the higher-order ($M>3$) orbitals are indispensable for correctly capturing the particle correlations, and indicate that more than one mode has contributions in each well. The single-band Bose-Hubbard model is an over-simplification for the system. 
We have seen that although the third orbital $\rho_3=4.432\times10^{-3}$ is two order of magnitude lower than the second orbital $\rho_2=0.493$, it has significant impact on the simulated particle correlations. 

In summary, to decide whether the convergence in orbital number has been achieved, we should perform simulations with more orbitals and compare the convergence in different quantities of interest.

\newpage

\begin{table}[h]
	\resizebox{\textwidth}{!}{
		\begin{tabular}{|l|l|c|c|c|c|c|c|}
			\hline
			Example N$^{\underline{o}}$ & & \textbf{\#1} & \textbf{\#2} & \textbf{\#3} & \textbf{\#4} & \textbf{\#5}& \textbf{\#6}\\
			\hline
			Use in the main text figures&  & \textbf{2(a-c)} & \textbf{2(d-f), 3} & \textbf{2(g-i), 3} & \textbf{2(j-l), 3} & \textbf{2(m-o), 3} & \textbf{3}\\
			\hline \hline
			Input variable names & \multicolumn{7}{|c|}{Basic parameters} \\ \hline
			\texttt{JOB\_TYPE} & Bosons or fermions & \texttt{BOS} & \texttt{BOS} & \texttt{BOS} & \texttt{FER} & \texttt{FER} & \texttt{BOS}\\
			\hline
			\texttt{Morb} & Number of orbitals $M$ & 10 & 10 & 10 & 10 & 10 & 8\\ 
			\hline
			\texttt{Npar} & Number of particles $N$ & 6 & 6 & 6 & 6 & 6 & 6\\
			\hline
			\texttt{Job\_Prefactor} & Relaxation or propagation & \multicolumn{6}{|c|}{(-1.0d0, 0.0d0)} \\
			\hline\hline
			& \multicolumn{7}{|c|}{One-body potential} \\ \hline
			\texttt{whichpot} & Potential $V(x)$ & \multicolumn{6}{|c|}{\textquotesingle\textquotesingle \texttt{HO1D+td\_gauss}\textquotesingle\textquotesingle}\\
			\hline
			\texttt{parameter1} & & 1.d0 & 1.d0 & 1.d0 & 1.d0 & 1.d0 & 1.d0\\
			\hline
			\texttt{parameter2} & & 0.d0 & 0.d0 & 0.d0 & 0.d0 & 0.d0 & 0.d0\\
			\hline
			\texttt{parameter3} & & 5.d0 & 20.d0 & 20.d0 & 20.d0 & 20.d0 & 20.d0\\
			\hline
			\texttt{parameter4} & & 0.01d0 & 0.01d0 & 0.01d0 & 0.01d0 & 0.01d0 & 0.01d0\\
			\hline
			\texttt{parameter5} & & 0.d0 & 0.d0 & 0.d0 & 0.d0 & 0.d0 & 0.d0\\
			\hline
			\texttt{parameter6} & & 0.5d0 & 0.5d0 & 0.5d0 & 0.5d0 & 0.5d0 & 0.5d0\\
			\hline\hline
			& \multicolumn{7}{|c|}{Two-body interaction} \\ \hline
			\texttt{xlambda\_0} & Interaction strength $g$ & 1.d0 & 1.d0 & 20.d0 & 0.d0 & 5.d0 & 0.d0\\
			\hline
			\texttt{which\_interaction} & Interaction $W(x,x')$ & \textquotesingle \texttt{gauss}\textquotesingle & \textquotesingle \texttt{gauss}\textquotesingle & \textquotesingle \texttt{gauss}\textquotesingle & \textquotesingle \texttt{regC}\textquotesingle & \textquotesingle \texttt{regC}\textquotesingle 
			& \textquotesingle \texttt{gauss}\textquotesingle\\
			\hline
			\texttt{Interaction\_Type} & Interaction $W(x,x')$& 0 & 0 & 0 & 4 & 4 & 0\\
			\hline
			\texttt{Interaction\_Parameter1} & & 0.d0 &  0.d0 & 0.d0 & 0.1d0 & 0.1d0 & 0.d0\\
			\hline
			\texttt{Interaction\_Parameter2} &  & 0.d0 & 0.d0 & 0.d0 & 100.d0 & 100.d0 & 0.d0\\
			\hline\hline
			& \multicolumn{7}{|c|}{Initial state, system and integration parameters} \\ \hline
			\texttt{GUESS} & Choose initial state & \multicolumn{6}{|c|}{\textquotesingle HAND\textquotesingle} \\
			\hline
			\texttt{Binary\_Start\_Time} & & \multicolumn{6}{|c|}{0.d0 (Not relevant)}\\
			\hline
			\texttt{DIM\_MCTDH} & System dimension & \multicolumn{6}{|c|}{1} \\
			\hline
			\texttt{NDVR\_X} & Number of grids & \multicolumn{6}{|c|}{512} \\
			\hline
			\texttt{x\_initial} & Grid left boundary &\multicolumn{6}{|c|}{-8.d0}\\
			\hline
			\texttt{x\_final} & Grid right boundary & \multicolumn{6}{|c|}{8.d0} \\
			\hline
			\texttt{Time\_Begin} & Initial time & \multicolumn{6}{|c|}{0.d0} \\
			\hline
			\texttt{Time\_Final}& Final time & 10.d0 & 10.d0 & 10.d0 & 10.d0 & 10.d0 & 30.d0\\
			\hline
			\texttt{Output\_TimeStep} & Output time step & \multicolumn{6}{|c|}{1.d0} \\
			\hline
			\texttt{Integration\_Stepsize} & Integration step size & \multicolumn{6}{|c|}{0.1d0} \\
			\hline
			\texttt{Coefficients\_Integrator} & & \multicolumn{6}{|c|}{\textquotesingle \texttt{DAV}\textquotesingle}\\
			\hline
		\end{tabular}
	}
	\caption{Important input variables in the input files for the relaxations shown in Figs.~2 and 3 of the main text. The inputs can be downloaded at \url{http://ultracold.org/data/tutorial_input_files.zip}. }
	\label{table:input_file_relax}
\end{table}

\begin{figure} [h]
\includegraphics[width=0.8\textwidth]{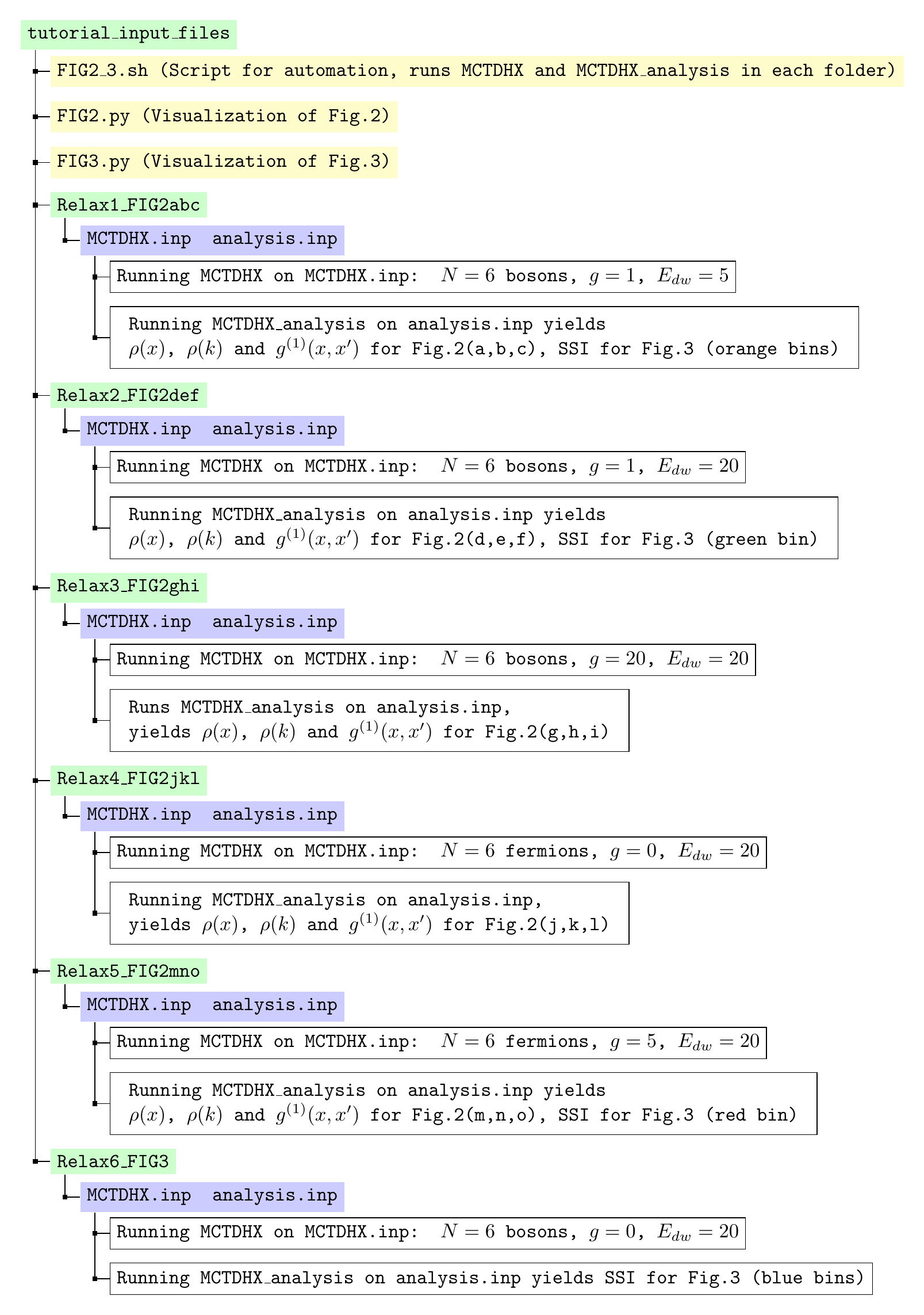}
	\caption{
		Files in \texttt{tutorial\_input\_files.zip} necessary to reproduce the results in Fig.~2 and Fig.~3 of the main text. By executing the command \texttt{./FIG2\_3.sh} (or \texttt{./FIG2\_3.sh X}, with \texttt{X} the number of parallel processes), all the results in Fig.~2 and Fig.~3 can be obtained and the figure can be plotted.
		The background colors indicate folders (green), input files (purple), executables (yellow), and explanations (white). Here, SSI is short for single-shot images. The input files can be downloaded at \htmladdnormallink{http://ultracold.org/data/tutorial\_input\_files.zip}{http://ultracold.org/data/tutorial_input_files.zip}.}
	\label{fig:tree1}
\end{figure}

\begin{table}[h]
	\centering
	\begin{tabular}{|l|c|c|c|c|}
		\hline
		Example N$^{\underline{o}}$ & \textbf{Series \#22} & \textbf{Series \#23} & \textbf{Series \#24} & \textbf{Series \#25} \\
		\hline
		Use in the main text figures&   \textbf{4(a), 5(f)} & \textbf{4(b)} & \textbf{4(c)} & \textbf{5(e)} \\
		\hline \hline
		Input variable names & \multicolumn{4}{|c|}{Basic parameters} \\ \hline
		\texttt{JOB\_TYPE} &  \texttt{BOS} & \texttt{BOS} & \texttt{FER} & \texttt{BOS} \\
		\hline
		\texttt{Morb} &  10 & 10 & 10 & 10 \\ 
		\hline
		\texttt{Npar} &  6 & 6 & 6 & 5 \\
		\hline
		\texttt{Job\_Prefactor} &  \multicolumn{4}{|c|}{(-1.0d0, 0.0d0)} \\
		\hline\hline
		& \multicolumn{4}{|c|}{One-body potential} \\ \hline
		\texttt{whichpot}  & \multicolumn{4}{|c|}{\textquotesingle\textquotesingle \texttt{HO1D+td\_gauss}\textquotesingle\textquotesingle}\\
		\hline
		\texttt{parameter1} & 1.d0 & 1.d0 & 1.d0 & 1.d0  \\
		\hline
		\texttt{parameter2} & 0.d0 & 0.d0 & 0.d0 & 0.d0 \\
		\hline
		\texttt{parameter3} & various & 20.d0 & 20.d0 & various \\
		\hline
		\texttt{parameter4} & 0.01d0 & 0.01d0 & 0.01d0 & 0.01d0 \\
		\hline
		\texttt{parameter5} & 0.d0 & 0.d0 & 0.d0 & 0.d0 \\
		\hline
		\texttt{parameter6} & 0.5d0 & 0.5d0 & 0.5d0 & 0.5d0 \\
		\hline\hline
		& \multicolumn{4}{|c|}{Two-body interaction} \\ \hline
		\texttt{xlambda\_0} & 1.d0 & various & various& 1.d0 \\
		\hline
		\texttt{which\_interaction}  & \textquotesingle \texttt{gauss}\textquotesingle & \textquotesingle \texttt{gauss}\textquotesingle & \textquotesingle \texttt{regC}\textquotesingle & \textquotesingle \texttt{gauss}\textquotesingle  \\
		\hline
		\texttt{Interaction\_Type} &  0 & 0 & 4 & 0\\
		\hline
		\texttt{Interaction\_Parameter1} & 0.d0 &  0.d0 & 0.1d0 & 0.d0\\
		\hline
		\texttt{Interaction\_Parameter2} & 0.d0 & 0.d0 & 100.d0 &  0.d0\\
		\hline\hline
		\multirow{2}{*}{ }& \multicolumn{4}{|c|}{Initial state, system} \\ 
		 & \multicolumn{4}{|c|}{and integration parameters} \\
		 \hline
		\texttt{GUESS} & \multicolumn{3}{|c|}{\textquotesingle HAND\textquotesingle} & \textquotesingle BINR\textquotesingle \\
		\hline
		\texttt{Binary\_Start\_Time} & \multicolumn{3}{|c|}{0.d0 (Not relevant)} & 10.d0\\
		\hline
		\texttt{Time\_Begin}  & \multicolumn{4}{|c|}{0.d0} \\
		\hline
		\texttt{Time\_Final} & \multicolumn{4}{|c|}{10.d0} \\
		\hline
		\texttt{Integration\_Stepsize} & \multicolumn{4}{|c|}{0.1d0} \\
		\hline
		\texttt{Coefficients\_Integrator} & \multicolumn{4}{|c|}{\textquotesingle \texttt{DAV}\textquotesingle}\\
		\hline
	\end{tabular}
	
	\caption{Important input variables in the input files for the relaxations shown in Fig.~4 and Fig.~5(e,f) of the main text. In the table, ``various'' means from 1.d0 to 20.d0 for \texttt{parameter3}, or from 0.d0 to 20.d0 for \texttt{xlambda\_0}. The inputs can be downloaded at \htmladdnormallink{http://ultracold.org/data/tutorial\_input\_files.zip}{http://ultracold.org/data/tutorial_input_files.zip}. }
	\label{table:input_file_series}
\end{table}

\begin{figure} [h]
	\includegraphics[width=0.9\textwidth]{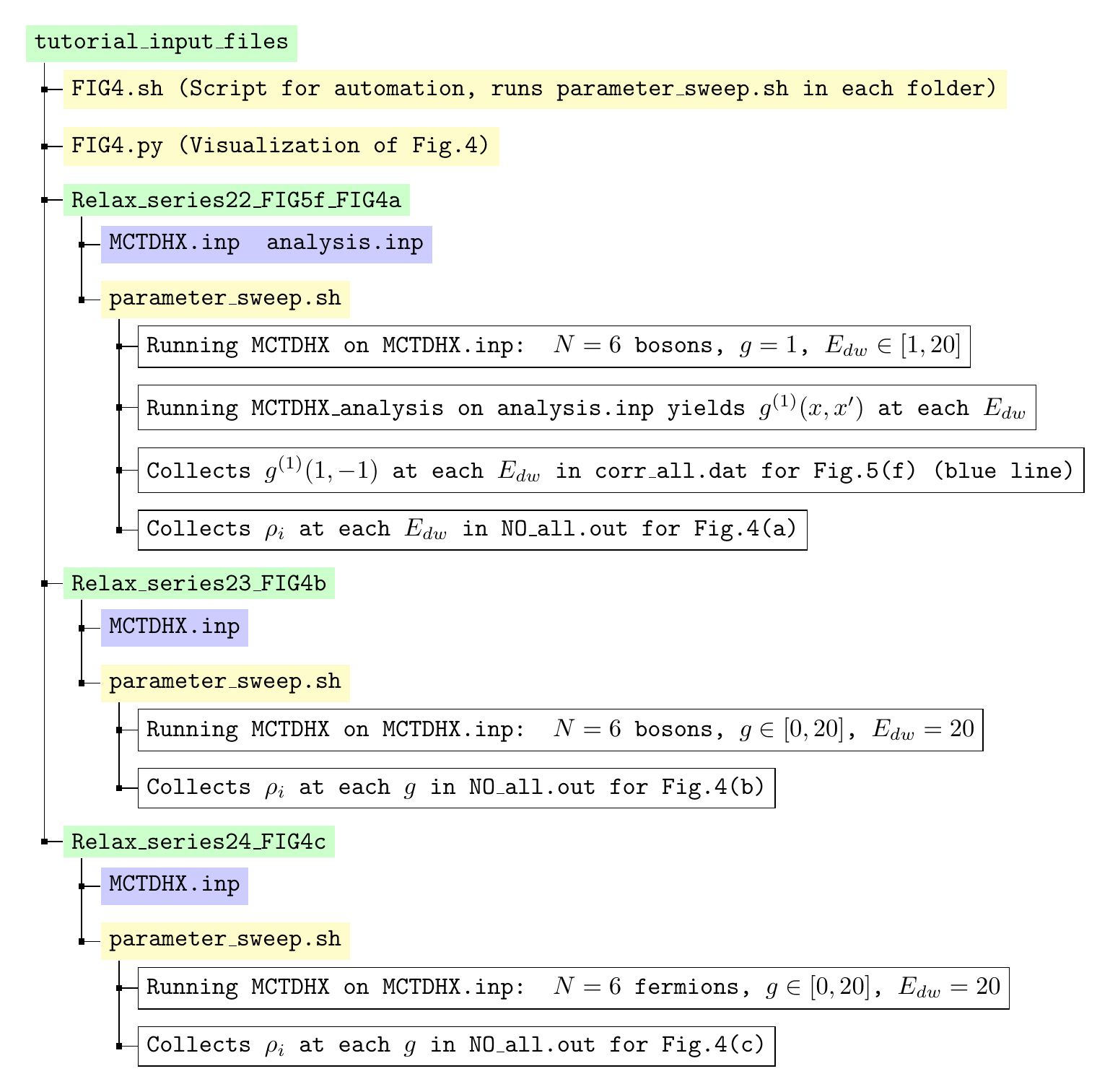}
	\caption{
	Files in \texttt{tutorial\_input\_files.zip} necessary to reproduce the results in Fig.~4 of the main text. By executing the command \texttt{./FIG4.sh} (or \texttt{./FIG4.sh X}, with \texttt{X} the number of parallel processes), all the results of Fig.~4 can be obtained and the figures can be plotted.
	The background colors indicate folders (green), input files (purple), executables (yellow), and explanations (white). The input files can be downloaded at \htmladdnormallink{http://ultracold.org/data/tutorial\_input\_files.zip}{http://ultracold.org/data/tutorial_input_files.zip}.}
	\label{fig:tree2}
\end{figure}

\begin{table}[h]
	\resizebox{\textwidth}{!}{
	\begin{tabular}{|l|c|c|c|c|c|c||c|c|c|c|c|}
		\hline
		Example N$^{\underline{o}}$ & \textbf{\#7} & \textbf{\#8} & \textbf{\#9} & \textbf{\#10} & \textbf{\#11} & \textbf{\# 12} & \textbf{\#13} & \textbf{\#14} & \textbf{\#15} & \textbf{\#16} & \textbf{\#17} \\
		\hline 
		Use in the main text figures &  & &  & \textbf{5(a)}  & & \textbf{5(c)} & \textbf{5(d)} & \textbf{5(b,e)} & \textbf{5(f)} & \textbf{5(e)} & \textbf{5(f)}\\
		\hline
		& \multicolumn{6}{|c||}{Relaxations for initial states} & \multicolumn{5}{|c|}{Propagations} \\
		\hline \hline
		Input variable names & \multicolumn{11}{|c|}{Basic parameters} \\ \hline
		\texttt{JOB\_TYPE} &  \texttt{FER} & \texttt{BOS} & \texttt{BOS} & \texttt{BOS} & \texttt{BOS} & \texttt{FER} &  \texttt{FER}  & \texttt{BOS} & \texttt{BOS} & \texttt{BOS} & \texttt{BOS} \\
		\hline
		\texttt{Morb} & 8 & 8 & 8 & 8 & 8 & 8 & 8 & 8 & 8 & 8 & 8\\ 
		\hline
		\texttt{Npar} & 5 & 5 & 6 & 5 & 6 & 5 & 5 & 5 & 6 & 5 & 6\\
		\hline
		\texttt{Job\_Prefactor} &  \multicolumn{6}{|c||}{(-1.0d0, 0.0d0)} &  \multicolumn{5}{|c|}{(0.0d0, -1.0d0)} \\
		\hline\hline
		& \multicolumn{11}{|c|}{One-body potential} \\ \hline
		\texttt{whichpot} &  \multicolumn{3}{|c|}{\textquotesingle\textquotesingle \texttt{HO1D}\textquotesingle\textquotesingle} & \multicolumn{6}{|c|}{\textquotesingle\textquotesingle \texttt{HO1D+td\_gauss}\textquotesingle\textquotesingle} & \multicolumn{2}{|c|}{\textquotesingle\textquotesingle \texttt{HO1D+td\_g\_r}\textquotesingle\textquotesingle}  \\
		\hline
		\texttt{parameter1} & 1.d0 & 1.d0 & 1.d0 & 1.d0 & 1.d0 & 1.d0 & 1.d0 & 1.d0 & 1.d0 & 1.d0 & 1.d0\\
		\hline
		\texttt{parameter2}  & 0.d0 & 0.d0 & 0.d0 & 0.d0 & 0.d0 & 0.d0 & 0.d0 & 0.d0 & 0.d0 & 0.d0 & 0.d0 \\
		\hline
		\texttt{parameter3}  & 0.d0 & 0.d0 & 0.d0 & 20.d0 & 20.d0 & 20.d0 & 20.d0 & 20.d0 & 20.d0 & 20.d0 & 20.d0\\
		\hline
		\texttt{parameter4}  & 0.d0 & 0.d0 & 0.d0 & 0.01d0 & 0.01d0 & 0.01d0 & 100.d0 & 100.d0 & 100.d0 & 100.d0 & 100.d0 \\
		\hline
		\texttt{parameter5}  & 0.d0 & 0.d0 & 0.d0 & 0.d0 & 0.d0 & 0.d0  & 0.d0 & 0.d0 & 0.d0 & 0.d0 & 0.d0\\
		\hline
		\texttt{parameter6}  & 0.d0 & 0.d0 & 0.d0 & 0.5d0 & 0.5d0  & 0.5d0 & 0.5d0 & 0.5d0 & 0.5d0 & 0.5d0  & 0.5d0\\
		\hline\hline
		& \multicolumn{11}{|c|}{Two-body interaction} \\ \hline
		\texttt{xlambda\_0}  & 2.d0 & 1.d0 & 1.d0 & 1.d0 & 1.d0  & 2.d0 & 2.d0 & 1.d0 & 1.d0 & 1.d0 & 1.d0\\
		\hline
		\texttt{which\_interaction}  & \textquotesingle \texttt{regC}\textquotesingle & \textquotesingle \texttt{gauss}\textquotesingle & \textquotesingle \texttt{gauss}\textquotesingle & \textquotesingle \texttt{gauss}\textquotesingle &\textquotesingle \texttt{gauss}\textquotesingle &\textquotesingle \texttt{regC}\textquotesingle & \textquotesingle \texttt{regC}\textquotesingle &\textquotesingle \texttt{gauss}\textquotesingle &\textquotesingle \texttt{gauss}\textquotesingle  &  \textquotesingle\texttt{gauss}\textquotesingle & \textquotesingle \texttt{gauss}\textquotesingle \\
		\hline
		\texttt{Interaction\_Type} & 4 & 0 & 0 & 0 & 0 & 4 & 4 & 0 & 0 & 0 & 0\\
		\hline
		\texttt{Interaction\_Parameter1} &  0.1d0 & 0.d0 & 0.d0 & 0.d0 & 0.d0 & 0.1d0 & 0.1d0 & 0.d0 & 0.d0 & 0.d0 &  0.d0\\
		\hline
		\texttt{Interaction\_Parameter2} & 100.d0 & 0.d0 & 0.d0 & 0.d0  & 0.d0  & 100.d0 & 100.d0 & 0.d0 & 0.d0 & 0.d0 & 0.d0\\
		\hline\hline
		& \multicolumn{11}{|c|}{Initial state, system and integration parameters} \\ \hline
		\texttt{GUESS} & \multicolumn{6}{|c||}{\textquotesingle \texttt{HAND}\textquotesingle} & \multicolumn{5}{|c|}{\textquotesingle \texttt{BINR}\textquotesingle} \\
		\hline
		\texttt{Binary\_Start\_Time} &  \multicolumn{6}{|c||}{0.d0 (Not relevant)} & \multicolumn{5}{|c|}{10.d0}\\
		\hline
		\texttt{Time\_Begin} & \multicolumn{6}{|c||}{0.d0} & \multicolumn{5}{|c|}{0.d0} \\
		\hline
		\texttt{Time\_Final} & \multicolumn{6}{|c||}{10.d0}  & \multicolumn{5}{|c|}{100.d0}\\
		\hline
		\texttt{Integration\_Stepsize} &  \multicolumn{6}{|c||}{0.1d0} & \multicolumn{5}{|c|}{0.01d0}  \\
		\hline
		\texttt{Coefficients\_Integrator} & \multicolumn{6}{|c||}{\textquotesingle \texttt{DAV}\textquotesingle} & \multicolumn{5}{|c|}{\textquotesingle \texttt{MCS}\textquotesingle}\\
		\hline	\hline
		& \multicolumn{11}{|c|}{Notes} \\ \hline
		Repetition$^*$ & & &  & Required & & &\multicolumn{5}{|c|}{N/A}\\
		\hline
		Binary file$^*$ & \multicolumn{6}{|c||}{N/A} & \#7 & \#8 & \#9 & \#10 & \#11\\
		\hline
	\end{tabular}
	}
	\caption{Important input variables in the input files for the results shown in Fig.~5 of the main text. The input variables \texttt{DIM\_MCTDH}, \texttt{NDVR\_X}, \texttt{x\_initial}, \texttt{x\_final}, \texttt{Output\_TimeStep} are unchanged from the Table~\ref{table:input_file_relax} and therefore not shown here. The six relaxations \#6 -- \#11 are used as initial states for the propagations \#12 -- \#17, respectively. The input files can be downloaded at \htmladdnormallink{http://ultracold.org/data/tutorial\_input\_files.zip}{http://ultracold.org/data/tutorial_input_files.zip}.}.
	\label{table:input_file_prop}
\end{table}

\begin{figure} [h]
	\includegraphics[width=0.9\textwidth]{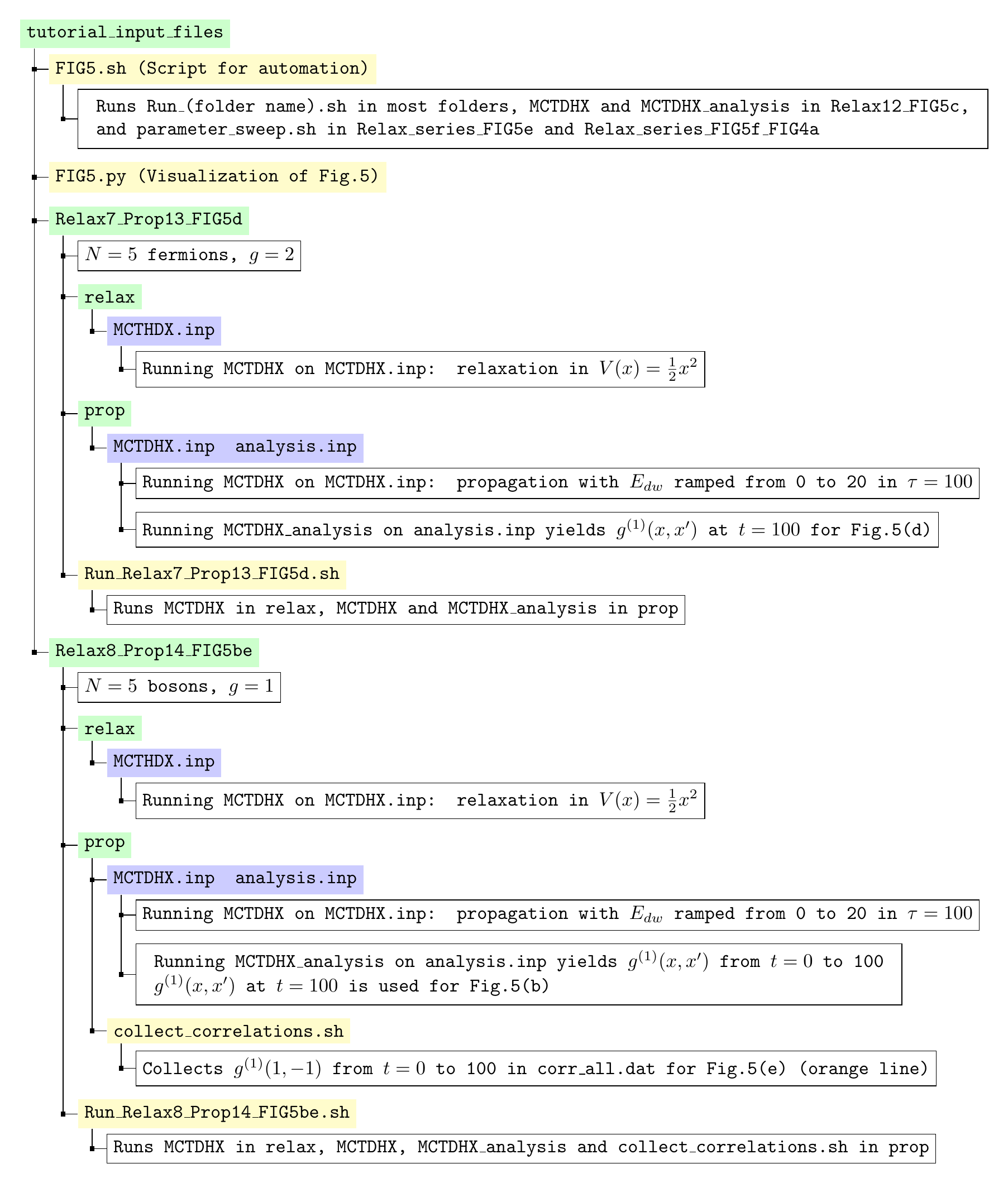}
	\caption{
	Files in \texttt{tutorial\_input\_files.zip} necessary to reproduce the results in Fig.~5 of the main text (part 1/3). By executing the command \texttt{./FIG5.sh} (or \texttt{./FIG5.sh X}, with \texttt{X} the number of parallel processes), all the results of Fig.~5 can be obtained and the figure can be plotted.
	The background colors indicate whether it is a folder (green), an input file (purple), an executable (yellow), or an explanation (white). Here, SSI is short for single-shot images. The input file \texttt{analysis2.inp} is used for the supplementary movie. The input files can be downloaded at \htmladdnormallink{http://ultracold.org/data/tutorial\_input\_files.zip}{http://ultracold.org/data/tutorial_input_files.zip}.}
	\label{fig:tree3_1}
\end{figure}

\begin{figure} [h]
	\includegraphics[width=0.9\textwidth]{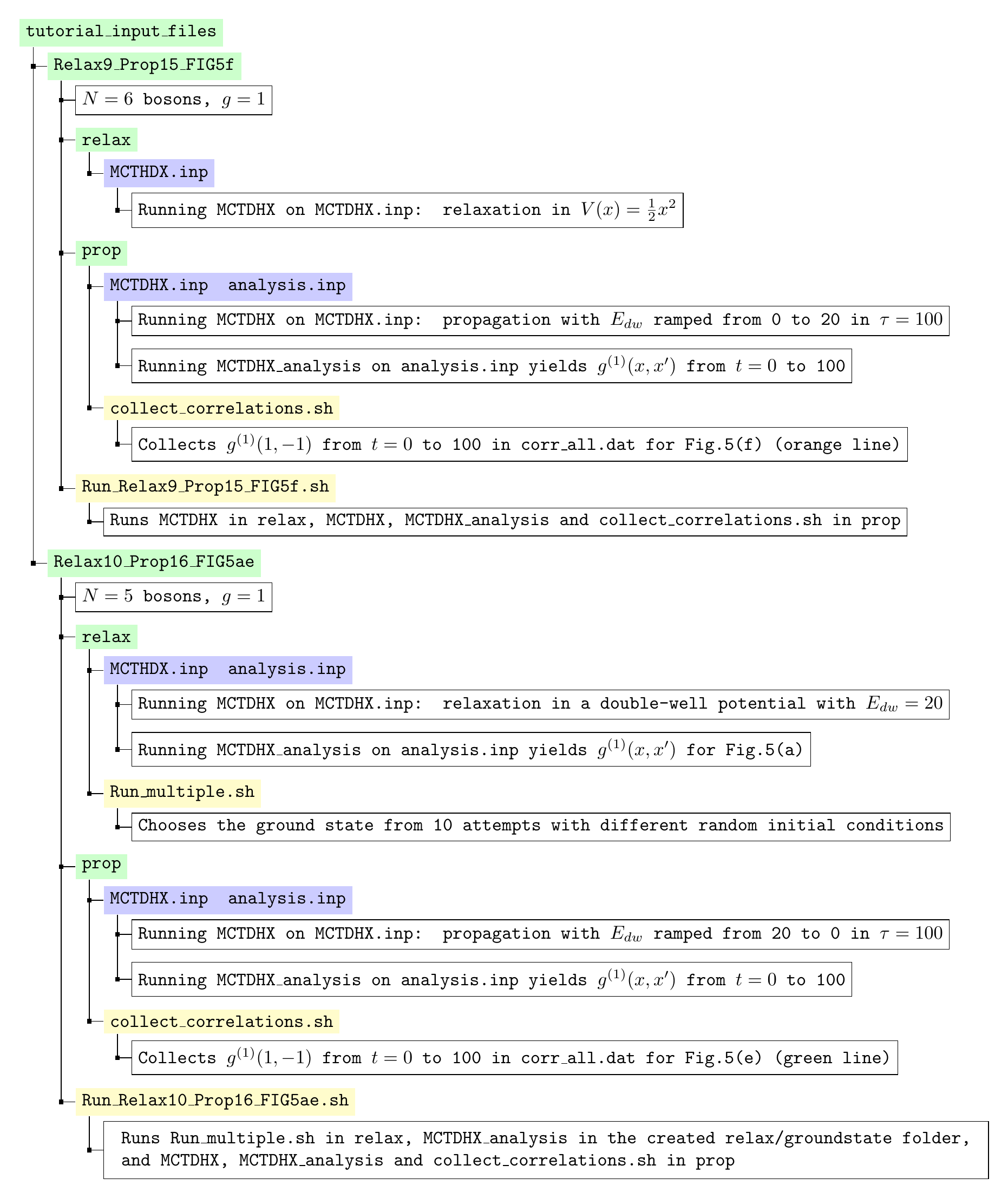}
	\caption{
		Files in \texttt{tutorial\_input\_files.zip} necessary to reproduce the results in Fig.~5 of the main text (part 2/3). By executing the command \texttt{./FIG5.sh} (or \texttt{./FIG5.sh X}, with \texttt{X} the number of parallel processes), all the results of Fig.~5 can be obtained and the figure can be plotted.
		The background colors indicate folders (green), input files (purple), executables (yellow), and explanations (white). Here, SSI is short for single-shot images. The input file \texttt{analysis2.inp} is used for the supplementary movie. The input files can be downloaded at \htmladdnormallink{http://ultracold.org/data/tutorial\_input\_files.zip}{http://ultracold.org/data/tutorial_input_files.zip}.}
	\label{fig:tree3_2}
\end{figure}

\begin{figure} [h]
	\includegraphics[width=0.9\textwidth]{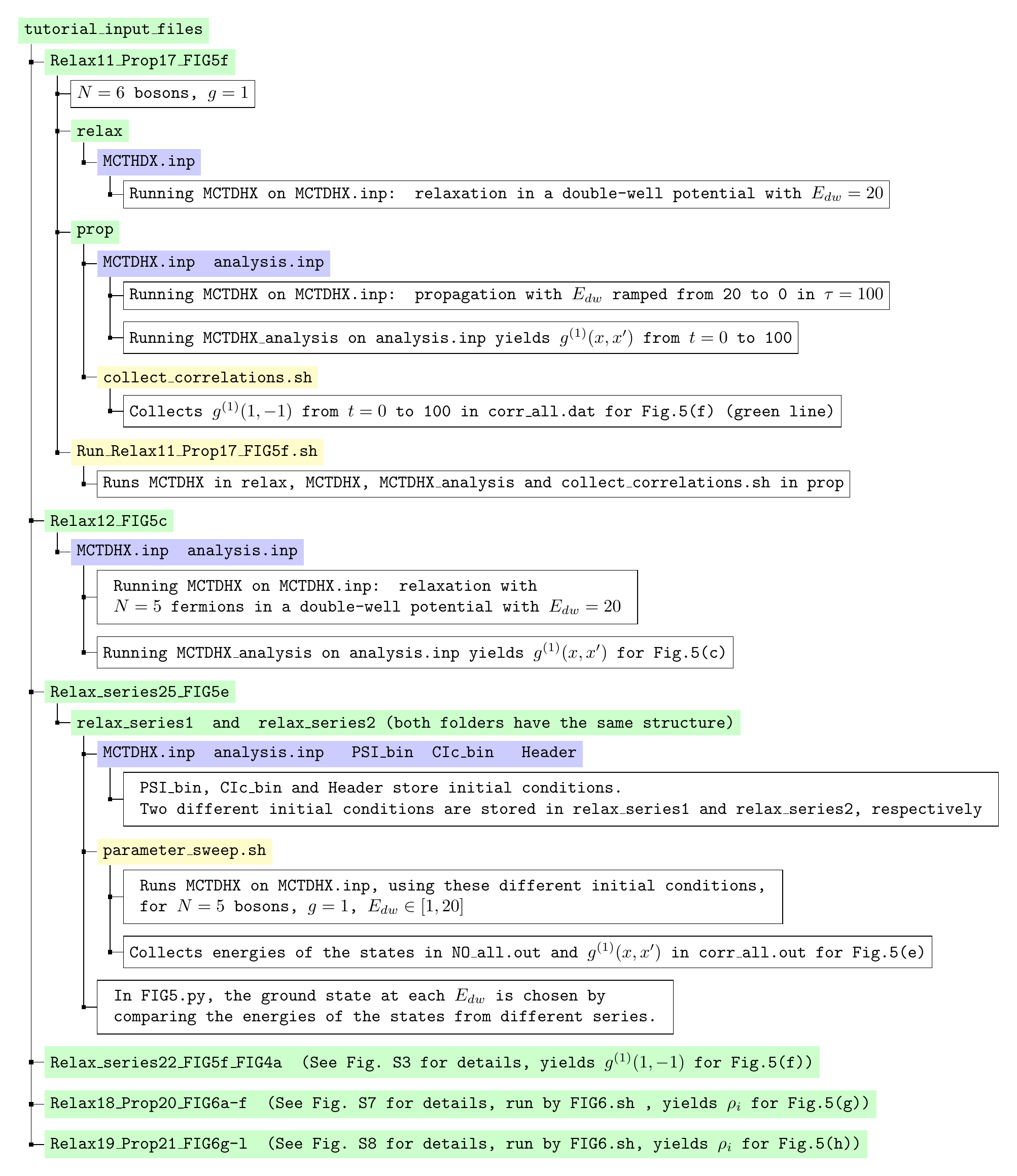}
	\caption{
		Files in \texttt{tutorial\_input\_files.zip} necessary to reproduce the results in Fig.~5 of the main text (part 3/3). By executing the command \texttt{./FIG5.sh} (or \texttt{./FIG5.sh X}, with \texttt{X} the number of parallel processes), all the results of Fig.~5 can be obtained and the figure can be plotted.
		The background colors indicate folders (green), input files (purple), executables (yellow), and explanations (white). Here, SSI is short for single-shot images. The input file \texttt{analysis2.inp} is used for the supplementary movie. The input files can be downloaded at \htmladdnormallink{http://ultracold.org/data/tutorial\_input\_files.zip}{http://ultracold.org/data/tutorial_input_files.zip}.
		}
	\label{fig:tree3_3}
\end{figure}

\begin{table}[h]
	\centering
	\begin{tabular}{|l|c|c||c|c|}
		\hline
		Example N$^{\underline{o}}$ & \textbf{\#18} & \textbf{\#19} & \textbf{\#20} & \textbf{\#21} \\
		\hline
		Use in the main text figures &  &  & \textbf{5(g), 6(a-f)} & \textbf{5(h), 6(g-l)} \\
		\hline
		& \multicolumn{2}{|c||}{Initial states} & \multicolumn{2}{|c||}{Propagations} \\
		\hline \hline
		Input variable names & \multicolumn{4}{|c|}{Basic parameters} \\ \hline
		\texttt{JOB\_TYPE} &  \texttt{BOS} & \texttt{FER} & \texttt{BOS} & \texttt{FER}  \\
		\hline
		\texttt{Morb} & 10 & 10 & 10 & 10 \\ 
		\hline
		\texttt{Npar} &  6 & 6 & 6 & 6\\
		\hline
		\texttt{Job\_Prefactor} &  \multicolumn{2}{|c||}{(-1.0d0, 0.0d0)} &  \multicolumn{2}{|c|}{(0.0d0, -1.0d0)} \\
		\hline\hline
		& \multicolumn{4}{|c|}{One-body potential} \\ \hline
		\texttt{whichpot} &  \multicolumn{2}{|c||}{\textquotesingle\textquotesingle \texttt{HO1D}\textquotesingle\textquotesingle} & \multicolumn{2}{|c|}{\textquotesingle\textquotesingle \texttt{HO1D+td\_gauss}\textquotesingle\textquotesingle}  \\
		\hline
		\texttt{parameter1} & 1.d0 & 1.d0 & 1.d0 & 1.d0 \\
		\hline
		\texttt{parameter2} & 0.d0 & 0.d0 & 0.d0 & 0.d0 \\
		\hline

		\texttt{parameter3} & 0.d0 & 0.d0 & 20.d0 & 20.d0 \\
		\hline
		\texttt{parameter4} & 0.d0 & 0.d0 & various & various  \\
		\hline
		\texttt{parameter5} & 0.d0 & 0.d0 & 0.d0 & 0.d0 \\
		\hline
		\texttt{parameter6} & 0.d0 & 0.d0 & 0.5d0 & 0.5d0 \\
		\hline\hline
		& \multicolumn{4}{|c|}{Two-body interaction} \\ \hline
		\texttt{xlambda\_0} & 1.d0 & 2.d0 & 1.d0 & 2.d0 \\
		\hline
		\texttt{which\_interaction}  & \textquotesingle \texttt{gauss}\textquotesingle & \textquotesingle \texttt{regC}\textquotesingle & \textquotesingle \texttt{gauss}\textquotesingle &  \textquotesingle \texttt{regC}\textquotesingle \\
		\hline 
		\texttt{Interaction\_Type} & 0 & 4 & 0 & 4 \\
		\hline
		\texttt{Interaction\_Parameter1} & 0.d0 & 0.1d0 & 0.d0 & 0.1d0 \\
		\hline
		\texttt{Interaction\_Parameter2} & 0.d0 & 100.d0 & 0.d0 & 100.d0 \\
		\hline\hline
		& \multicolumn{4}{|c|}{Initial state, system and integration parameters} \\ \hline
		\texttt{GUESS} & \multicolumn{2}{|c||}{\textquotesingle \texttt{HAND}\textquotesingle} & \multicolumn{2}{|c|}{\textquotesingle \texttt{BINR}\textquotesingle} \\
		\hline
		\texttt{Binary\_Start\_Time} &  \multicolumn{2}{|c||}{0.d0 (Not relevant)} & \multicolumn{2}{|c|}{10.d0}\\
		\hline
		\texttt{Time\_Begin} & \multicolumn{2}{|c||}{0.d0} & \multicolumn{2}{|c|}{0.d0} \\
		\hline
		\texttt{Time\_Final} & \multicolumn{2}{|c||}{10.d0}  & \multicolumn{2}{|c|}{100.d0}\\
		\hline
		\texttt{Integration\_Stepsize} &  \multicolumn{2}{|c||}{0.1d0} & \multicolumn{2}{|c|}{0.01d0}  \\
		\hline
		\texttt{Coefficients\_Integrator} & \multicolumn{2}{|c||}{\textquotesingle \texttt{DAV}\textquotesingle} & \multicolumn{2}{|c|}{\textquotesingle \texttt{MCS}\textquotesingle}\\
		\hline	\hline
		& \multicolumn{4}{|c|}{Notes} \\ \hline
		Binary file$^*$ & \multicolumn{2}{|c||}{N/A} & \#18 & \#19 \\
		\hline 
	\end{tabular}
	\caption{Important input variables in the input files for the results shown in Fig.~6 of the main text.  In the table, ``various'' means 1.d0, 10.d0, 50.d0 for different ramping rates. The input variables \texttt{DIM\_MCTDH}, \texttt{NDVR\_X}, \texttt{x\_initial}, \texttt{x\_final}, \texttt{Output\_TimeStep} are unchanged from the Table~\ref{table:input_file_relax} and therefore not shown here. The six  relaxations \#6 -- \#11 are used as initial states for the propagations \#12 -- \#17, respectively. The input files can be downloaded at \url{http://ultracold.org/data/tutorial_input_files.zip}.}.
	\label{table:input_file_prop2}
\end{table}

\begin{figure} [h]
	\includegraphics[width=0.9\textwidth]{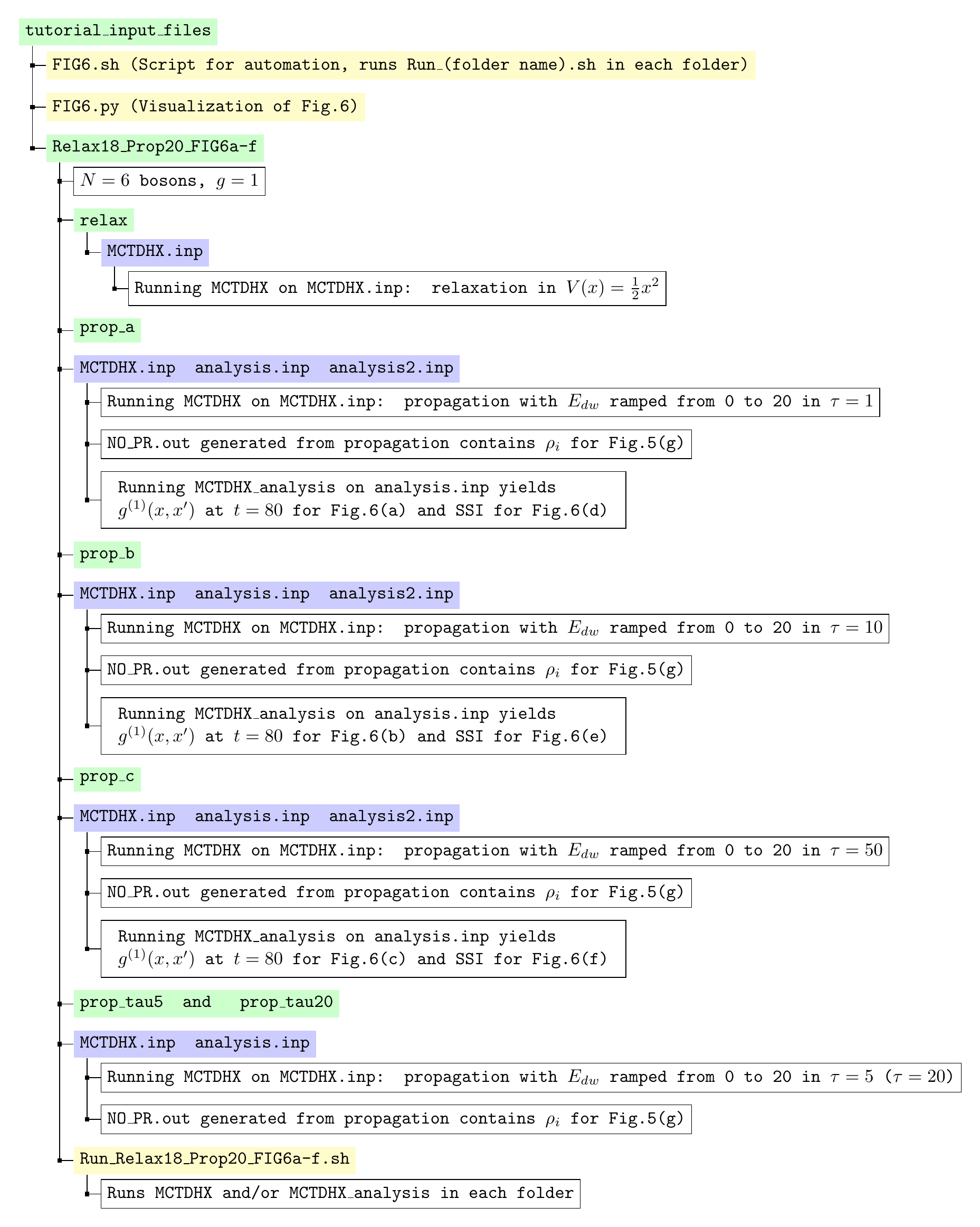}
	\caption{Files in \texttt{tutorial\_input\_files.zip} necessary to reproduce the results in Fig.~6 of the main text (part 1/2). By executing the command \texttt{./FIG6.sh} (or \texttt{./FIG6.sh X}, with \texttt{X} the number of parallel processes), all the results of Fig.~6 can be obtained and the figure can be plotted.
	The background colors indicate whether it is a folder (green), an input file (purple), an executable (yellow), or an explanation (white). Here, SSI is short for single-shot images. The input files \texttt{analysis2.inp} are used for generating $g^{(1)}(x,x')$ at $t=0\sim80$ for the supplementary movie. They should be renamed as \texttt{analysis.inp} when used. The input files can be downloaded at \htmladdnormallink{http://ultracold.org/data/tutorial\_input\_files.zip}{http://ultracold.org/data/tutorial_input_files.zip}.
	}
	\label{fig:tree4_1}
\end{figure}

\begin{figure} [h]
	\includegraphics[width=0.9\textwidth]{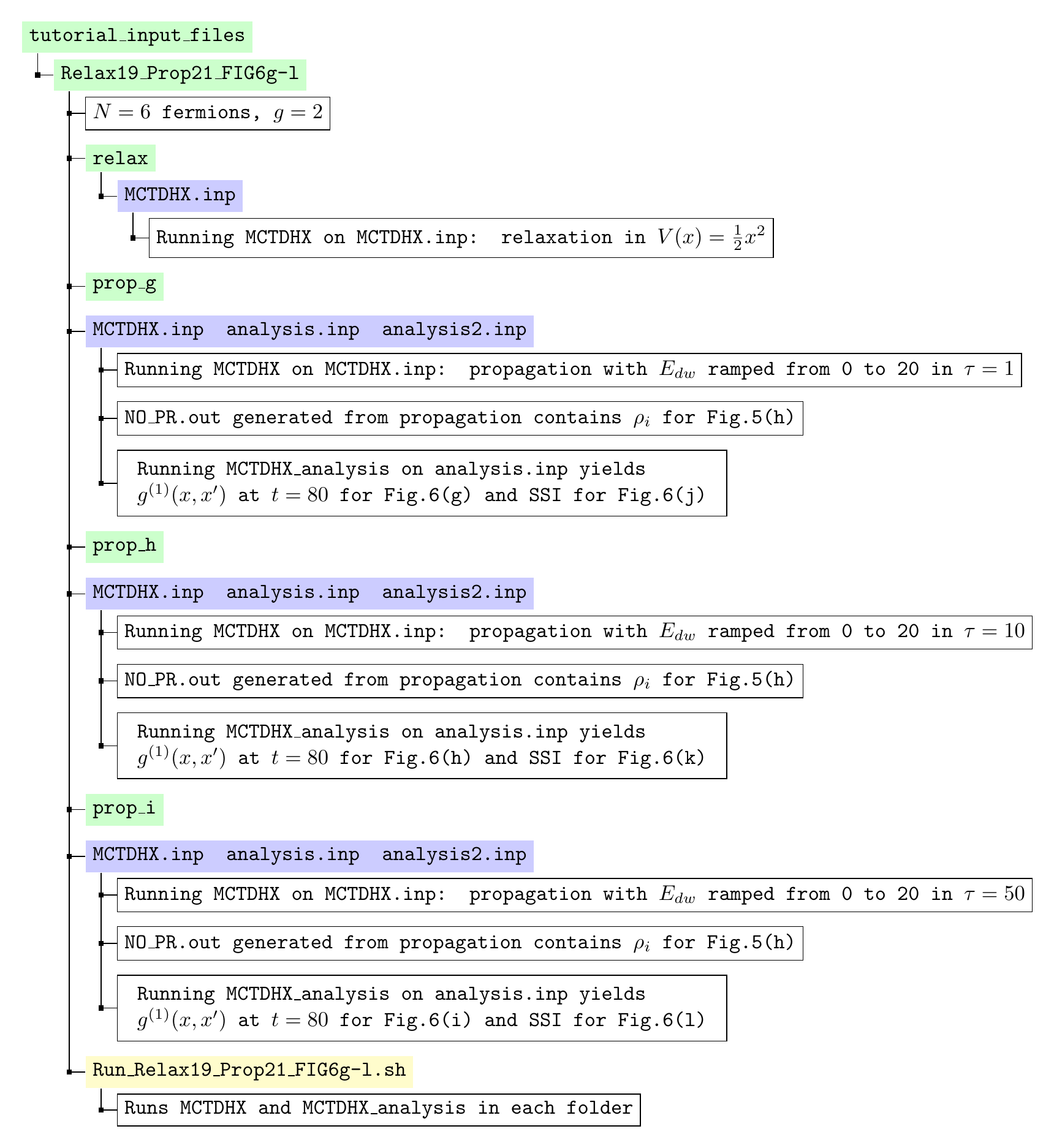}
	\caption{Files in \texttt{tutorial\_input\_files.zip} necessary to reproduce the results in Fig.~6 of the main text (part 2/2). By executing the command \texttt{./FIG6.sh} (or \texttt{./FIG6.sh X}, with \texttt{X} the number of parallel processes), all the results of Fig.~6 can be obtained and the figure can be plotted.
	The background colors indicate folders (green), input files (purple), executables (yellow), and explanations (white). Here, SSI is short for single-shot images. The input files \texttt{analysis2.inp} are used for generating $g^{(1)}(x,x')$ at $t=0\sim80$ for the supplementary movie. They should be renamed as \texttt{analysis.inp} when used. The input files can be downloaded at \htmladdnormallink{http://ultracold.org/data/tutorial\_input\_files.zip}{http://ultracold.org/data/tutorial_input_files.zip}.
		}
		\label{fig:tree4_2}
	\end{figure}

\begin{table}[h]
	\resizebox{\textwidth}{!}{
	\begin{tabular}{|l|l|c|c|c|}
		\hline
		& & For relaxations & For propagations & For propagations\\
		\hline
		Use in the simulations &   & (all relaxations) & \#13,\#20,\#21& \#14-\#17 \\
		\hline\hline
		Input variable names & \multicolumn{4}{|c|}{Zero body} \\ \hline
		\texttt{Time\_From} & Analysis starting time & 10.0d0 & 80.d0 or 100.d0 & 0.d0\\
		\hline
		\texttt{Time\_To} & Analysis final time & 10.0d0 & 80.d0 or 100.d0 & 100.d0 \\
		\hline
		\texttt{Time\_Points} & Number of analysis time point & 1 & 1 & 101 \\
		\hline
		\texttt{Dilation} & Resolution in the momentum space & 10 & 10 & 10\\
		\hline\hline
		& \multicolumn{4}{|c|}{One body} \\ \hline
		\texttt{Density\_x} & Real space density $\rho(x)$ & \texttt{.T.} & \texttt{.T.} & \texttt{.T.}  \\
		\hline
		\texttt{Density\_k} & Momentum space density $\tilde{\rho}(k)$ & \texttt{.T.} & \texttt{.T.} & \texttt{.T.} \\
		\hline
		\texttt{Correlations\_X} & One-body and two-body reduced density matrices & \texttt{.T.} & \texttt{.T.} & \texttt{.T.} \\
		\hline\hline
		& \multicolumn{4}{|c|}{Many body} \\ \hline
		\texttt{SingleShot\_Analysis} & Single shot analysis & \texttt{.T.} & \texttt{.T.} or \texttt{.F.} & \texttt{.F.} \\
		\hline
		\texttt{NShots} & Number of shots & 10000 & 10000 & 0 \\		
		\hline
	\end{tabular}
	}
	\caption{Important variables in the input files for the analysis used for Sec.~III in the main text. This \texttt{analysis.inp} extracts quantities of interest from the computed eigenstates and time-evolutions. The variable \texttt{SingleShot\_Analysis}  is set as \texttt{.F.} for simulation \#13, and \texttt{.T.} for simulations \#20 and \#21. The input files can be downloaded at \htmladdnormallink{http://ultracold.org/data/tutorial\_input\_files.zip}{http://ultracold.org/data/tutorial_input_files.zip}.}
	\label{table:input_file_analysis}
\end{table}

\begin{figure} [h]
	\includegraphics[width=\textwidth]{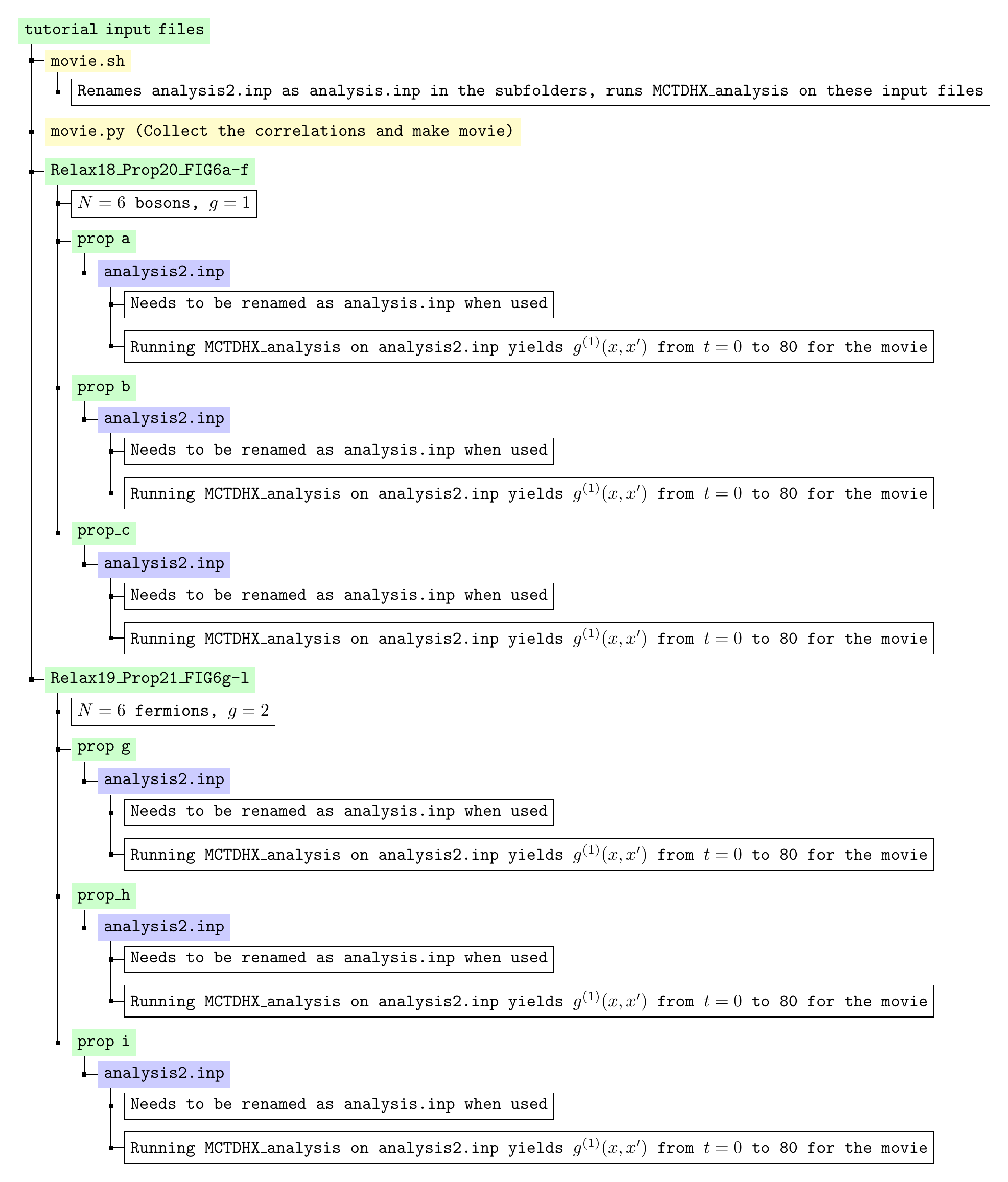}
	\caption{Files in \texttt{tutorial\_input\_files.zip} necessary to reproduce the movie on ~\htmladdnormallink{https://www.youtube.com/watch?v=l2UsTPmJ6po}{https://www.youtube.com/watch?v=l2UsTPmJ6po}. By executing the command \texttt{./movie.sh} after \texttt{./FIG6.sh}, the movie file can be generated.
	The background colors indicate folders (green), input files (purple), executables (yellow), and explanations (white). The input files can be downloaded at \htmladdnormallink{http://ultracold.org/data/tutorial\_input\_files.zip}{http://ultracold.org/data/tutorial_input_files.zip}.
	}
	\label{fig:tree5}
\end{figure}
\clearpage
\bibliography{MCTDHX_Mendeley}